\documentclass[12pt,a4paper]{article}
\usepackage{multirow}
\usepackage{authblk}
\usepackage{blindtext}
\usepackage{graphicx}
\usepackage{subcaption}
\usepackage{amsmath}
\usepackage{xfrac}
\usepackage{amssymb}
\newcommand{\be}{\begin{equation}} \newcommand{\ee}{\end{equation}}
\newcommand{\bea}{\begin{eqnarray}}\newcommand{\eea}{\end{eqnarray}}
\DeclareMathOperator{\sech}{sech}
\begin{document}

\title{Solvable Limits of a class of generalized
Vector Nonlocal Nonlinear Schr\"odinger equation with balanced loss-gain}
\author[]{Supriyo Ghosh\thanks{Email : supriyoghosh711@gmail.com} }
\author[]{Pijush K. Ghosh\thanks {Corresponding Author : pijushkanti.ghosh@visva-bharati.ac.in}}

\affil[]{Department of Physics, Siksha-Bhavana, Visva-Bharati, 
Santiniketan, PIN-731235, India.}
\date{\today}
\maketitle
\begin{abstract}
We consider a class of one dimensional Vector Nonlocal Non-linear 
Schr\"odinger Equation (VNNLSE) in an external complex potential
with time-modulated Balanced Loss-Gain(BLG) and Linear Coupling(LC) among 
the components of Schr\"odinger fields, and space-time dependent nonlinear 
strength. The system admits Lagrangian and Hamiltonian formulations under certain
conditions. It is shown that various dynamical variables like total power,
$\cal{PT}$-symmetric Hamiltonian, width of the wave-packet and its speed
of growth, etc. are real-valued despite the Hamiltonian density being complex-valued. 
We study the exact solvability of
the generic VNNLSE with or without a Hamiltonian formulation.
In the first part, we study time-evolution of moments which are analogous
to space-integrals of Stokes variables and find condition for existence of
solutions which are bounded in time. In the second part, we use a non-unitary
transformation followed by a coordinate transformation to map the VNNLSE to various
solvable equations. The coordinate transformation is not required at all for the
limiting case when non-unitary transformation reduces to pseudo-unitary
transformation.
The exact solutions are bounded in time for the same condition which
is obtained through the study of time-evolution of moments. 
Various exact solutions of the VNNLSE are presented.
\end{abstract}

\tableofcontents


\section{Introduction}

Nonlinear Schr\"odinger equation(NLSE) appears in many fields of 
science like optics in nonlinear media\cite{Kivshar,Serkin1,Serkin2}, Bose-Einstein condensates
(BEC)\cite{Dalfovo,Kevrekidis,Pitaevskii,Kengne}, gravity waves\cite{Trulsen}, plasma physics\cite{Dodd}, 
Bio-molecular dynamics\cite{Davydov} etc. Several generalisations of NLSE have been considered 
over the past few decades. One aspect of generalisations of NLSE is to introduce  non-local 
nonlinear interaction. Nonlocal nonlinearity arises whenever the nonlinear effect at a 
particular point depends on the influences from other points and it arises in many cases like 
Bose-Einstein condensation in a system having long range interaction\cite{Goral,Griesmaier}, 
transport process associated with heat conduction in media having a thermal influence, 
and during diffusion of charge carriers, atoms, or molecules in atomic vapours\cite{Tam,Suter}. 
In case of optical beams in nonlinear dielectric wave-guides or wave-guide arrays with random 
variation of refractive index size or wave-guide spacing, the effect of nonlocal nonlinearity 
becomes relevant\cite{Pertsch}. Nonlocal nonlinearities also arise in long-range 
interactions of molecules in nematic liquid crystals\cite{Conti}.

The exact solvability and integrability of non-local NLSE(NNLSE) have been 
studied in the last few years. Integrable NNLSE was introduced by Ablowitz 
and Musslimani\cite{Ablowitz1} and it admits both the bright and dark solitons for the 
same range of nonlinear strength\cite{Sarma}. The nonlocal vector NLSE is also integrable 
and soliton solution of the same equation is obtained through inverse scattering 
methods\cite{pkg1}. A discrete version of NNLSE has been shown to be exactly solvable and one
soliton solution has been constructed \cite{Ablowitz2}. Further, exact solutions of a class 
of NNLSE with confining potential like Harmonic trap, reflection-less potential and 
other $\cal{PT}$ symmetric complex potentials, have been obtained \cite{pkg2}. Various generalisations 
of the NNLSE having mathematical and physical importance have been 
considered\cite{Sarma,Agalarov,yang,Lakshmanan,Khare,Li,Huang,Ma1,Ma2,Wen,Chen1,Chen2,Liu,Tang,Sun, Saxena}. 
The dynamical behaviour of continuous and discrete NNLSE having $\cal{PT}$ invariant nonlinearity 
has been studied in Ref. \cite{Sarma}. It is shown in Ref. \cite{Agalarov} that NNLSE is gauge equivalent 
to physically important magnetic structure. For a number of nonlocal nonlinear equations, periodic and 
hyperbolic soliton solutions are reported in Ref. \cite{Khare}. Applying the N-th iterated Darboux 
transformation, a chain of non-singular localised wave solution which can describe 
the soliton interactions on continuous wave background, is derived for a NNLSE \cite{Li}. 
Localized nonlinear modes of self-focusing and defocusing NNLSE equation with 
$\cal{PT}$-symmetric complex potential like Rosen-Morse, Periodic, Scarf-II potential, 
are reported in Ref. \cite{Wen}.

Over the past few years, systems with balanced loss and gain(BLG) have received considerable
attention in the literature\cite{pkg-review,PRoy1,PRoy2}. Within this context, NLSE with BLG is known to admit 
bright\cite{Bludov1,Driben1} and dark solitons\cite{Bludov2}, breathers\cite{Barashenkov,Driben2}, 
rogue waves\cite{Kharif}, exceptional points\cite{Konotop,Driben2} etc. Soliton solution was found in 
the discrete non-linear Schr$\ddot{o}$dinger equation with gain and loss\cite{Suchkov}. 
Stabilisation of soliton by the periodic time-dependent BLG and LC parameters has 
been studied\cite{Driben2}. A specific model\cite{pkg3} with constant BLG and LC terms show power 
oscillations that may have interesting technological applications. Most of the papers dealing with 
the BLG system use approximation and numerical methods. A possible technique for obtaining 
exact solutions of coupled NLSE with arbitrary time dependent BLG parameters has been given in 
Ref.\cite{pkg4}. It has been shown in Ref.\cite{Sinha} that local and nonlocal vector NLSE with 
BLG is integrable when the matrix representing the linear term is pseudo-hermitian with respect 
to the hermitian matrix comprising the generic cubic nonlinearity.  

The article aims at identifying exactly solvable cases of VNNLSE in an external complex potential 
with BLG and LC among the components of Schr$\ddot{o}$dinger fields. The BLG and LC terms are allowed 
to be time dependent while the strength of nonlinear interactions is space-time dependent. The 
VNNLSE is a nonlocal generalization of the system discussed in Ref. \cite{pkg4} and 
may be used to model different physical phenomena wherein nonlocal interaction is present. 
We show that the  system admits Hamiltonian and Lagrangian formulations under certain conditions.
The subtleties involved in deriving Euler-Lagrange equations of motion and conserved Noether charges
associated with continuous internal symmetry for non-hermitian Lagrangian are discussed.
Further, we show  that various dynamical variables like charge, $\cal{PT}$-symmetric Hamiltonian, width
of the wave packets and its speed of growth, etc. are real valued despite the Hamiltonian density 
being complex valued.

We investigate the exact solvability of the generic VNNLSE with or without a Hamiltonian formulation. 
The study involves various combinations of time-dependent BLG, LC, external complex potential and 
space-time modulated nonlinear interaction. The time-evolution of certain moments, analogous to space-integrals 
of the familiar Stokes variables, is determined in terms of a set of coupled first order ordinary differential 
equations which is solved analytically. The method is an indirect way of studying collapse or growth of the 
wave-packet and has been used earlier\cite{pkg-wadati}. The exact solution of the system under investigation 
also allows to identify the regions in the parameter-space which admit bounded or unbounded solution in time. 
It should be noted that exact space-like dependence of the Schr\"odinger field can not be unearthed in this method. 
Nevertheless, it is an useful tool to study the bounded nature of the solution.

In the next step, we investigate the limits in which exact, analytical solution of the fields may be obtained. The
BLG and LC terms may be removed completely from the VNNLSE through non-unitary transformations provided the time-modulations of
the BLG and LC terms are identical\cite{pkg4}. In general, the non-unitary transformation modifies the nonlinear
term by imparting additional time-dependence. However, for the special case in which the nonlinear interaction may
be identified as the non-local analogue of the Manakov-Zakharov-Schulman(MZS) system, even the nonlinear interaction
remains unchanged. The non-unitary transformation reduces to the pseudo-unitary\cite{ali} transformation under
this situation.
It should be noted that this is not a symmetry transformation, since it does not keep the system invariant. Further,
unlike the unitary transformations, the non-unitary and pseudo-unitary transformations do not preserve the norm.
Thus, the removal of of the BLG and LC terms can not be thought of as gauge transformations. The mapping is used
to construct many exact solutions of the system for generic time-modulations of the BLG and LC terms and a large
class of complex external potentials.

We also investigate the system for which the non-unitary transformation removes the BLG and LC terms at the cost of 
modifying the nonlinear term. A co-ordinate transformation, as already discussed in ref. \cite{Kengne,pkg4,Beitia1}, is 
used to map the modified VNNLSE to solvable equation.
One essential condition for this co-ordinate transformation is that the transformed spatial co-ordinate is odd under 
parity transformation of original one. This condition is not applicable for its local counterpart
\cite{pkg4}. We present many exact solutions. 

The plan of the article is as follows. The model is introduced in Sec-2. The Lagrangian 
and Hamiltonian of the system are constructed in Sec.-2.1. The real valuedness of different 
dynamical variables like charge, Hamiltonian, width of wave packets etc. are shown. The study of 
the systems in terms of moments which are defined in terms of space-integrals of Stokes variables, are 
presented in the sec-2.2. In Sec-3, the VNNLSE is mapped to various solvable equations.
We present the solution for the case of a $M$-pseudo-hermitian $A$ in Sec. 3.1.
A few exact solutions for $V(x)=0$ are presented in Sec-3.1.1. The exact solutions for
a few non-vanishing complex confining potentials are presented in Sec-3.1.2. The general method for
which $A$ is not $M$-pseudo-hermitian is presented in Sec-3.2.  Finally the findings are summarized in Sec-4. 
In Appendix-I, several exact solutions for the case of pseudo-hermitian loss-gain matrix have been
presented.

\section{The Model}

We consider a VNNLSE with time-dependent BLG in terms of
a complex scalar field $\Psi(x,t) = (\psi_{1} \ \ \psi_{2})^{T}$ with
two components $\psi_{1}(x,t)$ and $\psi_{2}(x,t)$,
where the superscript ${}^{T}$ denotes the transpose of a matrix. The VNNLSE
with BLG has the form:
\begin{eqnarray}
i \psi_{1t} & = & -\psi_{1xx} + k^{*}(t) \psi_{2} + i \gamma(t) \psi_{1}
+ V(x,t) \psi_{1}\nonumber \\
	        & + & \sum_{i,j=1}^{2} \left [ l_{ij}(x,t) \ \psi_{i}^{*}(-x,t)\psi_{j}(x,t) \right ] \psi_1 \nonumber \\	
i \psi_{2t} & = & - \psi_{2xx} + k(t) \psi_{1} - i \gamma(t) \psi_{2} +
V(x,t) \psi_{2}\nonumber \\
	        & + & \sum_{i,j=1}^2 \left [ m_{ij}(x,t) \ \psi_{i}^{*}(-x,t)\psi_{j}(x,t) \right ] \ \psi_2	
\label{nlse1}
\end{eqnarray}
\noindent where $\psi_{it}$ and $\psi_{ixx}$ denote partial derivative of
$\psi_i$ with respect to $t$ and second order partial derivative of $\psi_i$
with respect to $x$, respectively. The time-dependent parameters $k(t)$ and
$\gamma(t)$ denote the strengths of the LC and the BLG, respectively. The 
space-time dependent real parameters $l_{ij}(x,t)$ and $m_{ij}(x,t)$ denote 
the strengths of the cubic nonlinearity. In the
context of optics, with $l_{11}$ and $m_{22}$ can be identified as
self-phase modulations, whereas $l_{12}$, $m_{11}$ can be identified as
cross-phase modulation. The external confining potential $V(x,t)$ appears in
the context of Bose-Einstein Condensates\cite{Pitaevskii}, which is taken to
be complex i.e. $V(x,t) = S(x,t) + i \tilde{S}(x,t)$ . There are many solvable
limits of Eq.(\ref{nlse1}), with the  simplest case $l_{ij} = m_{ij} = 0, \forall i,j$, which
models the optical wave propagation under paraxial approximation\cite{Ganainy,Markis}. 
With $M = L$ and vanishing LC, BLG and external potential, the system reduces to the nonlocal
analogue of  MZS system\cite{Stalin,Manakov,Zakharov}. 
A ${\cal PT}$-symmetry preserving breathing one-soliton solution has been constructed 
for the nonlocal Manakov equation through inverse scattering transform\cite{pkg1}.

We define an operator $D_{0} = \sigma_{0} \frac{\partial}{\partial t} + i A(t)$, where
the $2 \times 2$ matrix $A(t)$ is expressed as $A(t) = k \sigma_{-} + k^{*} \sigma_{+} + 
i \gamma \sigma_{3}$ and  $\sigma_{\pm} = \frac{1}{2}(\sigma_{1} \pm i \sigma_{2})$. $\sigma_0$ 
is a $2 \times 2$ identity matrix and $\sigma_{j}$'s, $j = 1,2,3$ are pauli matrices.
We also define the projection operator $P_{\pm}$ as $P_{\pm} = \frac{1}{2} (\sigma_{0} \pm \sigma_{3})$ 
with the property $P_{+}\Psi = (\psi_1 \ 0)^{T}$ and $P_{-} \Psi = (0 \ \psi_2)^{T}$. 
Eq. (\ref{nlse1}) can be written in a compact form by using matrix formulation, 
\begin{eqnarray}
i D_{0} \Psi & = & - \Psi_{xx} + V(x,t) \Psi+  [ \left \{ \Psi^{\dagger}(-x,t) L \Psi(x,t) \right \} P_{+} \nonumber \\
	     & + & \left \{ \Psi^{\dagger}(-x,t) M \Psi(x,t) \right \} P_{-} ] \  \Psi(x,t)
\label{nlse2}
\end{eqnarray}
\noindent where $\Psi^{\dagger}(-x,t) = \big(\psi_1^{*}(-x,t) \ \ \psi_2^{*}(-x,t) \big)$ and 
the expressions of matrix $M(x,t)$ and $L(x,t)$ are as follows,
\bea
L & = & l_{11} P_{+} + l_{22} P_{-} + l_{12} \sigma_{+} + l_{21} \sigma_{-} \nonumber \\
M & = & m_{11} P_{+} + m_{22} P_{-} + m_{12} \sigma_{+} + m_{21} \sigma_{-} 
\label{ML}
\eea
\noindent The form of the cubic nonlinearity of this VNNLSE is more general compared to 
its local counterpart i.e. Eq. (12) of Ref.\cite{pkg3}. With vanishing cross terms 
i.e. $l_{ij} = m_{ij} = 0$ ; $i \neq j$ form of the nonlinearity will be same for 
both the local and nonlocal NLSE. 

\subsection{Lagrangian, Hamiltonian formulations}

The non-local non-linear Schr\"odinger equation with or without balanced loss-gain terms
is described in terms of a non-relativistic field theory with complex-valued interaction.
Such systems with non-hermitian/complex interaction are of recent interests and belong to
the broad field of ${\cal{PT}}$-symmetric systems\cite{Millington1, Millington2, Millington3}.
Investigations in this field
are important towards having a full fledged ${\cal{PT}}$-symmetric quantum field theory.
The system admits a Lagrangian density in the limit $M = L$,
\bea
{\cal{L}} & = & \frac{i}{2} \left [ \Psi^{\dagger}(-x,t) M \big(D_0 \Psi(x,t)\big) - 
\big(D_0 \Psi(-x,t)\big)^{\dagger} M \Psi(x,t) \right ] \nonumber \\
          & - & \Psi_{x}^{\dagger}(-x,t)M \Psi_{x}(x,t) - \frac{1}{2} 
\left [ \Psi^{\dagger}(-x,t)M\Psi(x,t)\right ]^{2} \nonumber \\
	  & + & \frac{1}{2} \Psi^{\dagger}(-x,t) \left [ A^{\dagger}M - MA \right ]\Psi(x,t)-
\frac{i}{2} \Psi^{\dagger}(-x,t)M_{t}\Psi(x,t) \nonumber \\
          & - & \Psi^{\dagger}(-x,t)M_{x}\Psi_{x}(x,t) - V(x,t) \ \Psi^{\dagger}(-x,t) M \Psi(x,t) 
\eea
\noindent where $M_{t}$ and $M_{x}$ represent the differentiation of matrix $M$ with respect to time and space
respectively. Comparing with the Hamiltonian formulation of mechanical systems with BLG, the matrix $M$ may be 
interpreted as a background metric in which the system is defined\cite{pkg5,pkg6,pkg7,pkg8}. The field $\Psi(x,t)$ and its  
parity-transformed adjoint $\Psi^{\dagger}(-x,t)$ are treated
as two independent fields in contrast to the standard Lagrangian formulation of NLSE. The self-induced potential
$[\Psi^{\dagger}(-x,t) M \Psi(x,t)]^{2}$ in the corresponding stationary problem is non-hermitian, but,
${\cal{PT}}$-symmetric for a time-independent $M$ which is an even function of $x$.
The general prescription in higher dimensions in the context of ${\cal{PT}}$-symmetric non-local NLSE 
is to consider $\Psi(\vec{x},t)$ and $\Psi^{\dagger}({\cal{P}} \vec{x},t)$ as independent
fields\cite{pkg2}. In general, the Lagrangian density is non-hermitian and
${\cal{L}}^{\dagger}(-x,t) \neq {\cal{L}}(x,t)$, and care has to be taken in deriving  
Euler-Lagrange equations and conserved Noether charge associated with continuous symmetry
\cite{Millington1,Millington2,Millington3}. It should be noted that
${\cal{L}}^{\dagger}(-x,t) = {\cal{L}}(x,t)$ for a space-time independent $M$ and a
$M$-pseudo-hermitian $A$, i.e.  $A^{\dagger} = M A M^{-1}$ for which
the fourth, fifth and sixth terms in ${\cal{L}}$ vanish. The Lagrangian density ${\cal{L}}$
is still non-hermitian for this special case, however,  the property
${\cal{L}}^{\dagger}(-x,t) = {\cal{L}}(x,t)$ allows to derive the Euler-Lagrange equations
in a standard way.

The Euler-Lagrange equation w.r.t the variation 
of $\Psi^{\dagger}(-x,t)$ is obtained by  considering, 
\bea
\frac{\partial {\cal{L}}}{\partial \Psi^{\dagger}(-x,t)}-\frac{\partial}{\partial t}\left (\frac{\partial 
{\cal{L}}}{\partial \Psi^{\dagger}_{t}(-x,t)}\right) - \frac{\partial}{\partial x}
\left (\frac{\partial {\cal{L}}}{\partial \Psi^{\dagger}_{x}(-x,t)}\right) = 0
\label{el1}
\eea
\noindent which gives us 
\bea
i D_{0} \Psi & = & - \Psi_{xx} + V(x,t) \Psi + [ \Psi^{\dagger}(-x,t) M \Psi(x,t) ] \  \Psi(x,t)
\label{se1}
\eea
\noindent It may be noted that Eq. (\ref{nlse2}) with $L = M$ reduces to the above
equation. The Lagrangian density being non-hermitian and ${\cal{L}}^{\dagger}(-x,t)
\neq {\cal{L}}(x,t)$, the Euler-Lagrangian equation corresponding to the 
variation of $\Psi(x,t)$ is not equal to zero\cite{Millington1,Millington2,Millington3} i.e.
\bea
\frac{\partial {\cal{L}}}{\partial \Psi(x,t)}-\frac{\partial}{\partial t}
\big(\frac{\partial {\cal{L}}}{\partial \Psi_{t}(x,t)}\big) - \frac{\partial}{\partial x}
\big(\frac{\partial {\cal{L}}}{\partial \Psi_{x}(x,t)}\big) \neq 0
\label{el2}
\eea
\noindent The equation of motion of $\Psi^{\dagger}(-x,t)$ is obtained by complex 
conjugating Eq.(\ref{el1}) and then replacing $x$ by $-x$ i.e. from the following equation,
\bea
\frac{\partial {\cal{L}}^{\dagger}(-x,t)}{\partial \Psi(x,t)}-\frac{\partial}{\partial t}
\big(\frac{\partial {\cal{L}}^{\dagger}(-x,t)}{\partial \Psi_{t}(x,t)}\big) - \frac{\partial}{\partial x}
\big(\frac{\partial {\cal{L}}^{\dagger}(-x,t)}{\partial \Psi_{x}(x,t)}\big) = 0
\label{el3}
\eea
\noindent which gives,
\bea
i (D_0 \Psi(-x,t))^{\dagger} & = & \Psi^{\dagger}_{xx}(-x,t) - V^{*}(-x,t) \Psi^{\dagger}(-x,t) \nonumber \\                                           
                             & - & \big(\Psi^{\dagger}(-x,t)M^{\dagger}(-x,t)\Psi(x,t)\big) \Psi^{\dagger}(-x,t)
\label{se2} 
\eea
\noindent The above equation is the equation of motion of $\Psi^{\dagger}(-x,t)$. It may be noted
that equations (\ref{el2}) are  (\ref{el3}) are the same for
${\cal{L}}^{\dagger}(-x,t) = {\cal{L}}(x,t)$. It may be recalled that the non-local NLSE\cite{Ablowitz1}
or its higher dimensional\cite{pkg2} or multi-component generalization\cite{pkg1} can always be derived from
a Lagrangian density which is equal to its complex conjugate followed by $\vec{x} \rightarrow {\cal{P}} \vec{x}$.
The property ${\cal{L}}^{\dagger}(-x,t) = {\cal{L}}(x,t)$ is lost in presence of loss-gain terms without
the pseudo-hermitian property and/or space-time dependent metric $M$.

The Lagrangian 
density being non-hermitian, though it  has global $U(1)$ invariance i.e. it is invariant 
under the transformation $\Psi \rightarrow e^{-i\delta} \Psi$, there is no conserve 
current associated with this symmetry unlike the Noether's theorem 
for hermitian Lagrangian\cite{Millington1,Millington2,Millington3}. 
Considering the matrix $M$  hermitian and all it's elements are even functions of space and the potential 
is $\cal{PT}$-symmetric i.e. $V^{*}(-x,t) = V(x,t)$, we find that Eq. (\ref{se1}) and Eq. (\ref{se2}) admit
continuity equation. The expressions of charge density $\rho(x,t)$ 
and current density $J(x,t)$ are,
\bea
\rho(x,t) & = & \Psi^{\dagger}(-x,t) \eta \Psi(x,t) \nonumber \\
J(x,t)    & = & i \{\Psi^{\dagger}_x(-x,t) \eta \Psi(x,t) - \Psi^{\dagger}(-x,t) \eta \Psi_{x} \} 
\eea 
\noindent where $\eta$ is a constant matrix which appears in the definition of pseudo-hermitian 
$A(t)$ i.e. $A^{\dagger} = \eta A(t) \eta^{-1}$ \cite{ali}. 
We define a quantity $Q = \int \rho(x,t) \ dx $  as charge which is a conserved quantity. 
Though the charge density $\rho(x,t)$ is not real valued it can be shown that the charge $Q(x,t)$ 
is real valued by using the technique specified in Ref.\cite{pkg2}. We consider  
$\Psi(x,t) = \Psi_{e}(x,t) + \Psi_{o}(x,t)$,  where $\Psi_{e} = \frac{1}{2} \big(\Psi(x,t) 
+ \Psi(-x,t) \big)$ and $\Psi_{o} = \frac{1}{2} \big(\Psi(x,t) - \Psi(-x,t) \big)$. 
The subscripts 'e' and 'o' denote the even and odd functions respectively. 
The charge can be written as $\rho = \rho_r + \rho_c$ where $\rho_r$ and $\rho_c$ 
denote the real and imaginary parts of the charge density. The expressions of $\rho_r$ 
and $\rho_c$ can be expressed in terms of $\Psi_0$,$\Psi_e$ as shown below,
\bea
\rho_r & = & \big( \Psi^{\dagger}_e\eta\Psi_e - \Psi^{\dagger}_o\eta \Psi_e \big) \nonumber \\
\rho_c & = & \big( \Psi^{\dagger}_e\eta\Psi_o - \Psi^{\dagger}_o\eta \Psi_e \big)
\eea
\noindent The contribution coming from $\rho_c$ in $\int_{-\infty}^{\infty} \rho(x,t) dx$ is zero, 
since it is an odd function of $x$. Hence, the charge $Q(x,t)$ is real valued. The momentum corresponding 
to $\Psi(x,t)$ and $\Psi^{\dagger}(-x,t)$ are $\Pi_{\Psi(x,t)} = \frac{i}{2} \Psi^{\dagger}(-x,t)M$ 
and $\Pi_{\Psi^{\dagger}(-x,t)} = - \frac{i}{2} M \Psi$. The Hamiltonian density of the system is given as,
\bea
{\cal{H}} & = & \Psi^{\dagger}(-x,t) M A \ \Psi(x,t) + \Psi_{x}^{\dagger}(-x,t)M \Psi_{x} + 
\frac{1}{2} \big(\Psi^{\dagger}(-x,t)M\Psi(x,t)\big)^{2} \nonumber \\
          & + & \frac{i}{2} \Psi^{\dagger}(-x,t)M_{t}\Psi(x,t) + \Psi^{\dagger}(-x,t)M_{x}\Psi_{x}(x,t) \nonumber \\
          & + & V(x,t) \ \Psi^{\dagger}(-x,t) \Psi(x,t) 
\eea
\noindent Though the Hamiltonian density ${\cal{H}}$ is not real-valued, we can show that the Hamiltonian i.e. 
$H = \int {\cal{H}} dx $ is real valued when $M$ is considered as constant Hermitian matrix and 
$A(t)$ is $M$-pseudo-hermitian i.e. $A^{\dagger} = M A M^{-1}$. The Hamiltonian density ${\cal{H}}$ can be written 
as ${\cal{H}} = {\cal{H}}_{r} + {\cal{H}}_{c}$, where ${\cal{H}}_{r}$ is real and ${\cal{H}}_{c}$ 
is purely imaginary. The expressions of ${\cal{H}}_{r}$ and ${\cal{H}}_{c}$ are given below, 
\bea
{\cal{H}}_{r} & = & \big(\Psi^{\dagger}_e MA \Psi_e - \Psi^{\dagger}_o M A \Psi_o \big) + 
\big( \Psi^{\dagger}_e MA \Psi_o - \Psi^{\dagger}_o MA \Psi_e \big) \nonumber \\
              & + & \big( \Psi^{\dagger}_{ex} M \Psi_{ex} - \Psi^{\dagger}_{ox} M \Psi_{ox} \big) 
+ \frac{1}{2} \big(\Psi^{\dagger}_e M \Psi_e - \Psi^{\dagger}_o M \Psi_o \big)^{2} \nonumber \\
              & + & \frac{1}{2} \big(\Psi^{\dagger}_e M \Psi_o - \Psi^{\dagger}_o M \Psi_e \big)^{2} 
+ S(x,t) \big( \Psi^{\dagger}_e \Psi_e - \Psi^{\dagger}_o \Psi_{o} \big) \nonumber \\
              & + & i \ \tilde{S}(x,t) \big( \Psi^{\dagger}_e \Psi_o - \psi^{\dagger}_o \Psi_e \big)  \nonumber 
\eea
\bea
{\cal{H}}_{c} & = & \{ \big(\Psi^{\dagger}_{ex} M \Psi_{ox} - \Psi^{\dagger}_{ox} M \Psi_{ex} \big) \} 
+ \frac{1}{2} \big \{ \big(\Psi^{\dagger}_{e} M \Psi_{e} - \Psi^{\dagger}_{o} M \Psi_{o} \big) 
\big(\Psi^{\dagger}_{e} M \Psi_{o} - \Psi^{\dagger}_{o} M \Psi_{e} \big) \nonumber \\
             & + & \big(\Psi^{\dagger}_{e} M \Psi_{o} - \Psi^{\dagger}_{o} M \Psi_{e} \big) 
\big(\Psi^{\dagger}_{e} M \Psi_{e} - \Psi^{\dagger}_{o} M \Psi_{o} \big) \big \} + 
S(x,t) \big(\Psi^{\dagger}_{e} \Psi_{o} - \Psi^{\dagger}_{o}\Psi_{e} \big) \nonumber \\
             & + & i \tilde{S}(x,t) \big( \Psi^{\dagger}_{e} \Psi_{e} - \Psi^{\dagger}_{o} \Psi_{o} \big)  
\eea 
\noindent It is to be noted that when the potential is $\cal{PT}$ symmetric, i.e. $V^{*}(-x,t) = V(x,t)$
then ${\cal{H}}_{c}$ is an odd function of x. Hence it will not contribute to the  
Hamiltonian $H = \int_{-\infty}^{\infty} {\cal{H}} dx $. 
Similarly it can be shown that $\int x^{2n} \rho(x,t)$ is real valued where n is a positive integer. 
The moment $\int x^{2} \rho(x,t) dx$ is identified as width of the wave packets. We define growth 
speed of the system as $\int x J(x,t) dx$ which is also real valued. The expression of 
current density $J(x,t)$ in terms of $\Psi_{e}(x,t)$,$\Psi_{o}(x,t)$ can be written as,
\bea
J(x,t)     & = & J_{c} + J_{r} \nonumber \\
J_{r}(x,t) & = & i\big( \Psi_{ex}^{\dagger}\eta\Psi_{e} - \Psi_{e}^{\dagger}\eta \Psi_{ex}
- \Psi_{ox}^{\dagger}\eta\Psi_{o} + \Psi_{o}^{\dagger}\eta\Psi_{ox} \big) \nonumber \\
J_{c}(x,t) & = & i \big( \Psi_{ex}^{\dagger}\eta\Psi_{o}+\Psi_{o}^{\dagger}\eta \Psi_{ex}
- \Psi_{ox}^{\dagger}\eta\Psi_{e}-\Psi_{e}^{\dagger}\eta\Psi_{ox} \big) \nonumber \\
\eea
\noindent $J_{r}$ and $J_{c}$ are the real and imaginary parts of current density. Now
the real part $J_r$ being an  odd function of $x$, $\int_{-\infty}^{\infty} J(x,t) dx$, receives 
contribution from imaginary part of $J$. Hence the current will be imaginary. But in 
case of speed of the growth of the system i.e. $\int_{-\infty}^{\infty} x J(x,t) dx$, 
the imaginary part of $x J(x,t)$ being an odd function of $x$, will vanish and the real 
part will survive. Hence Speed of the growth of the system is real valued.

\subsection{Dynamics of Moments}

A large number of nonlinear equations are not amenable to exact solutions. The moment
method is an indirect way for studying the stabity of a nonlinear equation without actually
solving it\cite{pkg-wadati}. On the other hand, pseudo-hermitian and pseudo-unitary transformation are
extensively used in the context of quantum mechanics\cite{ali}. We combine these two apparently
disjoint methods to study the stability of the nonlinear equation with non-local and non-hermitian
interaction. It will be seen that a direct outcome of the method is the controlling of instabilities,
if any, in a systematic way through suitable choice of time-modulation of linear coupling and loss-gain terms.
The new method is no less important than finding exact solutions, and is expected to be applicable to
a wide variety of nonlinear equations.

The time-evolution of the system defined by Eq. (\ref{se1}) may be studied
in terms of the moments defined as,
\bea
Z_{a}(t) = \int dx \Psi^{\dagger}(-x,t)\eta \sigma_a \Psi(x,t), \ a = 0, 1, 2, 3
\eea
\noindent The method is applicable for field configurations for which 
$\Psi(x,t), \Psi_x(x,t)$ are continuous on the whole line $x$ and vanishes
in the limit ${\vert x \vert} \rightarrow \infty$. Further, closed form
expressions for the equations governing the time-evolution of the moments 
are possible only if $A$ is $\eta$-pseudo-hermitian, i.e. $A^{\dagger}= \eta
A \eta^{-1}$. It follows from Eqs. (\ref{se1}) and (\ref{se2}) that, for
the specific field-configurations stated above and an $\eta$-pseudo hermitian
$A$, the moments $Z_{a}$ satisfy the following equations:
\bea
\dot{Z_{a}} = i \int \Psi^{\dagger}(-x,t) \eta \left [A(t), \sigma_{a} \right ]
\Psi(x,t) dx \ + \ \int \Psi^{\dagger}(-x,t) \eta_t \sigma_a \Psi(x,t) dx 
\label{m1} 
\eea
\noindent Here, $Z_0$ is the conserved charge $Q$ and $Z_{j}$ , $j = 1,2,3$ are analogues of
spatial integration of Stokes variables. The expression of $\eta(t)$ for  $\eta(t)$-pseudo-hermitian 
$A(t) = [k^{*}(t) \sigma_{+} + k(t) \sigma_{-} + i \gamma(t) \sigma_{3} ] $ is given as\cite{pkg3},
\bea
\eta(t) = \frac{|\tilde{\alpha}| |k(t)|}{\gamma(t)} \sigma_0 \sin(\theta_{\tilde{\alpha}} 
- \theta_k) + \tilde{\alpha}^{*} \sigma_{+} + \tilde{\alpha} \sigma_{-}
\label{metric1}
\eea
\noindent where $k(t) = |k(t)| e^{i\theta_k(t)}$ and $\tilde{\alpha} = |\tilde{\alpha}| e^{i\theta_{\tilde{\alpha}}}$ 
is an arbitrary constant complex number. The expression reduces to Eq.(\ref{metric2}) 
for $k(t) = \mu_0(t) \ \beta$ and $\gamma(t) = \mu_0(t) \ \Gamma$. The co-efficient of $\sigma_0$  is the 
only time dependent term in $\eta(t)$ and we denote this term as $h(t) = \frac{|\tilde{\alpha}| 
|k(t)|}{\gamma(t)} \sin(\theta_{\tilde{\alpha}}-\theta_k)$. We define two quantities $a(t)$, $b(t)$ as 
\bea
a(t) = 2 |k| sin(\theta_k) + \frac{i |\tilde{\alpha}| sin(\theta_{\tilde{\alpha}}) h_t }{|\eta|} \ , 
\ b(t) =  2 |k| cos(\theta_k) + \frac{i |\tilde{\alpha}| cos(\theta_{\tilde{\alpha}}) h_t}{|\eta |}
\eea 
\noindent where $|\eta|$ is the determinant of $\eta$. 
The Eq.(\ref{m1}) can be transformed into a set of coupled linear differential equations as,
\bea
\dot{Z} = N(t) Z, \ 
Z = \begin{pmatrix}
Z_0 \\
Z_1 \\
Z_2 \\
Z_3 
\end{pmatrix} \ , \
 N(t) = \begin{bmatrix}
\frac{h_t h}{|\eta |} & -\frac{\alpha_1 h_t}{|\eta|} & -\frac{\alpha_2 h_t}{|\eta|} & 0 \\
-\frac{\alpha_1 h_t}{|\eta|} & \frac{h_t h}{|\eta |} & -2i\gamma(t) & a \\
-\frac{\alpha_2 h_t}{|\eta|} & 2i\gamma & \frac{h_t h}{|\eta |} & -b(t) \\
0 & -a(t) & b(t) & \frac{h_t h}{|\eta |}
\end{bmatrix}
\label{m2}
\eea 
\noindent The Eq.(\ref{m2}) is not solvable for arbitrary time dependence of the parameters $k(t)$ 
and  $\gamma(t)$. If $N(t)$ commutes with $\int^{t} N(t') dt'$ i.e. 
$\Big[ N(t) \ , \ \int^{t} N(t') dt' \Big] = 0$, Eq. (\ref{m2}) leads to the solution 
$Z = e^{\int^{t} N(t') dt'} \ C$, where $C = \Big( C_0 \ C_1 \ C_2 \ C_3 \Big)^{T}$ is a 
constant column matrix. We will see that this condition is also necessary for obtaining 
exact solution using the general methods discussed in Sec-3. We choose 
$k(t) = \mu_0(t) \beta$ and $\gamma(t) = \mu_0(t) \Gamma$ where 
$\beta \in \mathbb{C}$, \ $\Gamma \in \mathbb{R}$ and $\mu_0(t)$ is an arbitrary real function.  
The commutation relation, $\Big[ N(t) \ , \ \int^{t} N(t') dt' \Big] = 0$ is satisfied 
with these choices. Further, $h_{t} = 0$ and  $\eta$ will be time 
independent and $Z_0$ which is actually the modified power is a constant of motion. 
Now excluding constant $Z_0$ term Eq.(\ref{m2}) can be written as follows,
\bea
\dot{Z} = \mu_0(t) N_0 Z \ , \ 
Z = \begin{pmatrix}
Z_1 \\
Z_2 \\
Z_3 
\end{pmatrix} \ , \
N_0 = \begin{bmatrix}
0 & -2i\Gamma & 2\beta_2 \\
2i\Gamma & 0 & -2\beta_1 \\
-2\beta_2 & 2\beta_1 & 0
\end{bmatrix}
\label{m3}
\eea
\noindent where $\beta_1$ and $\beta_2$ are the real and imaginary parts of $\beta$. 
Eigen value of matrix $N_0$ are $0$,\ $\pm 2i\sqrt{|\beta|^2-\Gamma^2}$. 
Evaluating $e^{\int^{t} \mu_0(t') N_0}$, the expressions of $Z_1$, $Z_2$, $Z_3$ are given as follows,
\bea
Z_1 & = & [-2 \beta_1 {\vert \beta \vert}^{2} (\beta_1 C_1 + \beta_2 C_2 + i \Gamma C_3) \nonumber \\
    & + & 2 \{ C_1 (\Gamma^2 \beta_1^2 - \epsilon_0^2 \beta_2^2) + C_2 \beta_1 \beta_2 |\beta|^2 
+ iC_3|\beta|^2\beta_1 \Gamma \} \cos(2\epsilon) \nonumber \\
    & + & 2 \epsilon_0 {\vert \beta \vert}^{2} ( i C_2 \Gamma - C_3 \beta_2) \sin(2\epsilon)] \nonumber \\
Z_2 & = & [-2 \beta_2 |\beta|^2 \ \{\beta_1 C_1 + \beta_2 C_2 + i \Gamma C_3\} \nonumber \\
    & + & 2 \{ C_1 |\beta|^2 \beta_1\beta_2 + C_2 (\Gamma^2 \beta_2^2 - \epsilon_0^2 \beta_1^2) 
+ iC_3|\beta|^2\beta_2\Gamma \} \cos(2\epsilon)  \nonumber \\
    & + & 2 \epsilon_0 |\beta|^2 \ \{ -iC_1\Gamma + C_3\beta_1 \}] \nonumber \\
Z_3 & = & [-2i\Gamma |\beta|^2 \{\beta_1 C_1 + \beta_2 C_2 + i \Gamma C_3 \} \nonumber \\
    & + & 2 |\beta|^2 \{ iC_1\Gamma\beta_1 + iC_2\Gamma\beta_2-C_3|\beta|^2 \} \cos(2\epsilon) \nonumber \\
    & + & 2 \epsilon_0 |\beta|^2 \{\beta_2 C_1 - \beta_1 C_2 \} \sin(2\epsilon)] 
\label{ex-z1}
\eea  
\noindent We can also solve Eq. (\ref{m3}) by transforming $Z \rightarrow \tilde{Z}$ as 
$\tilde{Z} = \tilde{P}^{-1} Z$ such that $\tilde{P}$ is the diagonalizing matrix of matrix $N_0$.
\bea
\tilde{P} = \begin{pmatrix}
\beta_1 & \Gamma \beta_1 + \epsilon_0 \beta_2 & \Gamma \beta_1 - \epsilon_0 \beta_2 \\
\beta_2 & \Gamma \beta_2 - \epsilon_0 \beta_1 & \Gamma \beta_2 + \epsilon_0 \beta_1 \\
i \Gamma & i|\beta|^2 & i |\beta|^2 
\end{pmatrix}
\eea
\noindent We get the expressions of $Z_{j}$'s as shown below,
\bea
Z_1 & = & \beta_1 \tilde{C}_1 + (\Gamma \beta_1 + \epsilon \beta_2) \tilde{C}_2 e^{2i\epsilon} 
+ (\Gamma \beta_1 - \epsilon \beta_2) \tilde{C}_3 e^{-2i\epsilon} \nonumber \\
Z_2 & = &  \beta_2 \tilde{C}_1 + (\Gamma \beta_2 -\epsilon \beta_1) \tilde{C}_2 e^{2i\epsilon} 
+ (\Gamma \beta_2 + \epsilon \beta_1) \tilde{C}_3 e^{-2i\epsilon} \nonumber \\
Z_3 & = & i \Gamma \tilde{C}_1 + i |\beta|^2 \tilde{C}_2 e^{2i\epsilon} 
+ i |\beta|^2 \tilde{C}_3 e^{-2i\epsilon} \nonumber \\
\label{sl_m}
\eea  
\noindent As shown in the first part of Sec-3, $\epsilon = \epsilon_0 \mu(t)$. It may seem 
that Eq. (\ref{ex-z1}) and Eq. (\ref{sl_m}) are different. We can write 
$\tilde{C_1},\tilde{C_2},\tilde{C_3}$ in terms of $C_1,C_2,C_3$ and $\beta$, $\Gamma$ so 
that both the expressions of $Z_1,Z_2,Z_3$ are same. Both of these 
methods are applicable when matrix $\tilde{P}$ is invertible i.e. $det(\tilde{P}) \neq 0$ which implies 
$|\beta|^2 \neq \Gamma^2$ and $ |\beta| \neq 0$. For constant BLG and LC terms, 
the moments $Z_1$,$Z_2$,$Z_3$ show periodic nature for the real $\epsilon_0$ 
i.e. $\vert \beta \vert ^{2} > \Gamma^2$. Otherwise the solutions blow up. But with 
the appropriate choice of $\mu(t)$, as for example $\mu(t) = sin(t)$ the moments  show periodic 
behaviour even for $\vert \beta \vert ^{2} < \Gamma^2$. This condition is consistent with the 
condition of finite solution of Eq.(\ref{nlse1}). We can choose constants $\tilde{C}_1,
\tilde{C}_2,\tilde{C}_3$ appropriately. For $\tilde{C}_1 = 0$, $\tilde{C}_3 = -\tilde{C}_2$ and 
$\tilde{C}_2 \in \mathbb{R}$, Eq.(\ref{sl_m}) can be written as,
\bea
Z_1 & = & \tilde{C}_2 [ 2i \Gamma \beta_1 \sin(2\epsilon) + 2 \epsilon \beta_2 \cos(2\epsilon)] \nonumber \\
Z_2 & = & \tilde{C}_2 [ 2i\Gamma \beta_2 \sin(2\epsilon) - 2 \epsilon \beta_1 \cos(2\epsilon)] \nonumber \\
Z_3 & = & - 2 |\beta |^2 \tilde{C}_2 \sin(2\epsilon)
\eea
\noindent where $Z_3$ is real. Both the real and imaginary parts of 
$Z_1$, $Z_2$ are periodic in time. Besides the conserved charge, we can construct 
two other constants of motion as, 
\bea
Q_2 & = & \beta_1 Z_1 + \beta_2 Z_2 + i \Gamma Z_3 \nonumber \\
Q_3 & = & Z_1^2 + Z_2^2 + Z_3^2 \nonumber
\eea
\noindent After transforming $Z \rightarrow \tilde{Z}$ it is found that $\dot{\tilde{Z}} = 0$ which 
gives the conserved quantity $Q_2$. The matrix $N_0$ being skew-symmetric matrix, from 
Eq. (\ref{m3}) we get that the time derivative $Z^{T}Z$ is zero i.e. $Q_3$ is constant of motion.

\section{Transformation to solvable equations}

We remove the balanced loss gain term by transforming the fields $\Psi(x,t)$ 
into $\Phi(x,t)$ via non unitary transformation as follows,
\begin{eqnarray}
\Psi(x,t) = U(t) \Phi(x,t)
\label{unitary1}
\end{eqnarray}
\noindent where $U(t)$ is chosen so that the equation $i U_{t} - A U = 0$ holds leading to 
the solutions $U(t) = e^{-i \int^{t} A(t) dt}$ for $[A(t),\int^{t} A(t) dt ] = 0 $\cite{pkg4}. $U(t)$
maps $\Phi(x,t)$ to $\Psi(x,t)$ at the same time t. It should not be confused with time evolution 
operator. Since $A(t)$ is non-hermitian, $U(t)$ is non-unitary.
We consider $k(t) = \mu_0(t) \beta$ and $\gamma(t) = \Gamma \mu_0(t)$ where $\mu_0(t)$, $\Gamma$ are real 
parameters. We introduce few functions as $\mu(t) = \int^{t} \mu_0(t) dt$, $\epsilon_0 = \sqrt{|\beta|^2 - \Gamma^2}$ 
and $\epsilon = \epsilon_0 \mu(t)$ for which $A(t)$ can be written as $A(t) = \mu_0(t) A_0$ where 
$A_0 = \beta^{*} \sigma_{+} + \beta \sigma_{-} + i \Gamma \sigma_3$. Now $U(t)$ can be expressed as,

\begin{eqnarray}
U(t) & = & \sigma_0 \cos(\epsilon) - \frac{i A_0}{\epsilon_0} \sin(\epsilon)
\label{punitary}  
\end{eqnarray}
\noindent In the limit of $\epsilon_0$ is zero and purely imaginary, the expressions of $U(t)$ 
becomes as follows,
\bea
U(t) & = & \sigma_0 - i A_0 \mu(t) \ \  (for \ \epsilon_0 = 0)  \nonumber \\
U(t) & = & \cosh(|\epsilon_0|\mu(t)) - i \frac{A_0}{|\epsilon_0|} \sinh(|\epsilon_0|\mu(t)) \ \ 
(for \ \epsilon_0 = i |\epsilon_0|)
\eea
\noindent When the LC parameter $k$ and BLG parameter $\gamma$ are constant, $\epsilon_0$ must be real
i.e. $|\beta|^{2} > \Gamma^2$. Otherwise, $U(t)$ will grow with time and the solution will be unbounded. 
But for the suitable choice of $\mu_0(t)$, bounded solution is possible for $\epsilon_0$ is zero and 
purely imaginary. For example, with the choice
\bea
\mu_0(t) = \cos(t),
\eea
\noindent $\mu(t)$ reduces to $\sin(t)$ and U(t) corresponding to all the three cases discussed 
above are periodic in time.
Using the transformation as shown in Eq.(\ref{unitary1}), Eq.(\ref{nlse2}) reduces to the following,
\begin{eqnarray}
i \Phi_{t}(x,t) & = & - \Phi_{xx}(x,t) + V(x,t) \Phi(x,t) +
[ K(l_{11},m_{11}) \Phi^{\dagger}(-x,t) F_{+} \Phi(x,t) \nonumber \\
& + & K(l_{22},m_{22}) \Phi^{\dagger}(-x,t) F_{-} \Phi(x,t) \nonumber \\
& + & K(l_{12},m_{12}) \Phi^{\dagger}(-x,t) B \Phi(x,t) \nonumber \\
& + & K(l_{21},m_{21}) \Phi^{\dagger}(-x,t) B^{\dagger} \Phi(x,t) ] \Phi(x,t)
\label{nlse3}
\end{eqnarray}
\noindent where $K(\zeta_1,\zeta_2) = \zeta_1 U^{-1} P_{+} U + \zeta_2 U^{-1}
P_{-} U $, $F_{\pm} = U^{\dagger} P_{\pm} U$ and $B = U^{\dagger} \sigma_{+} U$.
The operator $K$ takes a simple form for $\zeta_1=\zeta_2$, in particular,
$K(\zeta_1, \zeta_1)=\sigma_0$. This is also the limit for a hermitian $K$. Form of 
the cubic nonlinearity of Eq. (\ref{nlse3}) is more general compared to its local counterpart 
i.e. Eq. (12) of Ref. \cite{pkg4}. 
The operators $F_{\pm}$ and $K(\zeta_1, \zeta_2)$ have the same expressions 
for both local as well as Non-local NLSE. The explicit
expressions of $F_{\pm}$ and $K(\zeta_1, \zeta_2)$ have been obtained for the
first time in Ref.\cite{pkg4} in the context of local NLSE and these are reproduced
in this article for completeness. With the introduction of the function $T_{\pm}$ 
\begin{eqnarray}
T_{\pm} & = & \frac{\Gamma \mu(t)}{\epsilon^2} \sin^2(\epsilon) \pm
\frac{\sin(2\epsilon)}{2\epsilon}
\end{eqnarray}
\noindent the explicit expressions of $K(\zeta_1,\zeta_2)$ and $F_{\pm}$ are, 
\begin{eqnarray}
&& K(\zeta_1,\zeta_2)= \frac{\zeta_1+\zeta_2}{2} \sigma_0 + i (\zeta_1 - \zeta_2)
\mu(t) \big(\beta^{*} T_{-} \sigma_{+} + \beta T_{+} \sigma_{-} \big) \nonumber \\
&&+ (\zeta_1 - \zeta_2) (\frac{1}{2} - \frac{\mu^2 |\beta|^{2}}{\epsilon^2}
\sin^{2}(\epsilon) \big) \sigma_3,\nonumber \\
&& F_{\pm} = \big( \frac{1}{2} + \Gamma \mu(t) T_{\pm} \big) \sigma_0 -
i T_{\pm} \mu(t) (\beta^{*} \sigma_{+} - \beta \sigma_{-}) \nonumber \\
&& + \big( \frac{\Gamma \mu(t)}{2\epsilon} \sin(2 \epsilon) \pm \frac{1}{2}
\cos(2 \epsilon) \big) \sigma_3 
\end{eqnarray}
\noindent The operator $B$ arises due to the cross-phase modulation terms
and has the explicit expression,
\begin{eqnarray}
B & = & \sigma_{+} \cos^{2}(\epsilon) - \frac{i \beta}{2 \epsilon_0} \sigma_3 \sin(2\epsilon) 
+ (\beta^2 \sigma_{-} - \Gamma^2 \sigma_{+} - i \Gamma \beta \sigma_0 ) \frac{\sin^2(\epsilon)}{{\epsilon_0}^2} 
\end{eqnarray}
\noindent The operator $B$ is non-hermitian.

At this point, it is important to consider whether or not the solution of Eq. (\ref{nlse2}) that 
corresponds to a stable solution of Eq. (\ref{nlse3}) is likewise stable. The transformation 
(\ref{unitary1}) that joins the initial system given by Eq. (\ref{nlse2}) to Eq. (\ref{nlse3}),  
does not change the stability property of $\Phi(x,t)$ for a bounded U(t). In other words, the solution 
$\Psi(x,t) = U(t) \ \Phi(x,t)$ is a stable solution for a bounded $U(t)$ provided $\Phi(x,t)$ is stable 
under perturbation. This is demonstrated by taking into 
account the equation $\Phi(x,t) = \tilde{\Phi}(x,t) + \chi(x,t)$, where $\tilde{\Phi}(x,t)$ is the exact 
solution to equation (\ref{nlse3}) and $\chi(x,t)$ is a tiny perturbation. When we enter the expression 
for $\Phi(x,t)$ into equation (\ref{nlse2}), we get the equation of motion of $\chi(x,t)$. 
The stability of the exact solution is determined by Eq. of motion of $\chi(x,t)$ and the solution is 
stable if $\chi$ is a bound state. The exact solution of Eq, (\ref{nlse2}) corrresponding to $\tilde{\Phi}$ 
is $\tilde{\Psi} = U \tilde{\Phi}$. We can analyse the stability of Eq. (\ref{nlse2}) by perturbing the exact 
solution $\tilde{\Psi}$. We take the purturbation term as $ U \eta(x, t)$, where U (t) is 
be bounded in time and $\eta(x,t)$ is an arbitrary small fluctuations. We find after plugging $\Psi(x,t) = 
\tilde{\Psi}(x,t) + U \eta(x,t)$ into Eq. (\ref{nlse2})  that $\eta(x,t)$ satisfies the same Eq. of motion 
as that of $\chi(x,t)$ . Thus, the same Eq. dictates the stability of both the original Eq. (\ref{nlse2}) 
and the transformed Eq. (\ref{nlse3}). Hence, the transformation (\ref{unitary1}) preserves the
stability property for a bounded U in time and identical initial conditions.

\subsection{$M$-Pseudo-hermitian $A$}

In the limit $L = M$ and vanishing BLG, LC and external potentials Eq. (\ref{nlse2}) models 
the nonlocal analogue of MZS system\cite{Manakov,Zakharov}. MZS system with constant BLG and LC 
terms have been studied in Ref. \cite{pkg3}. In this subsection we will consider nonlocal analogue 
of MZS system along with time dependent BLG and LC terms. Eq. (\ref{nlse2}) admits Lagrangian 
formulations for $L = M$.
The limiting case $L=M$ deserves a special attention for which Eq. (\ref{nlse3})
takes a simpler form, 
\bea
i \Phi_{t}(x,t) = - \Phi_{xx}(x,t) + V(x,t) \Phi(x,t) +
\left  [ \Phi^{\dagger}(-x,t) U^{\dagger} M U \Phi(x,t) \right ] \Phi(x,t)
\label{p-herm1}
\eea 
\noindent  The effect of the non-unitary transformation is to remove the
BLG terms at the cost of modifying the term
$\Phi^{\dagger}(-x,t) M \Phi(x,t) \Rightarrow \Phi^{\dagger}(-x,t) U^{\dagger}
M U \Phi(x,t)$. If we further assume that $A(t)$ is $M$-pseudo hermitian, i.e.
$A^{\dagger}=M A M ^{-1}$, then $U(t)$ becomes pseudo-unitary, i.e. $U^{\dagger}M U
= M$. Thus, for a $M$-pseudo hermitian $A$, Eq. (\ref{p-herm1}) reduces to 
the non-local analogue of the standard vector NLSE in an external potential
$V(x,t)$:
\bea
i \Phi_{t}(x,t) = - \Phi_{xx}(x,t) + V(x,t) \Phi(x,t) +
\left  [ \Phi^{\dagger}(-x,t) M \Phi(x,t) \right ] \Phi(x,t)
\label{p-herm2}
\eea 
\noindent It may be noted that the pseudo-unitary transformation is not
a symmetry transformation of the system \textemdash it removes the BLG
terms and keeps all other interaction terms invariant. The technique has been
used for the first time in the context of local NLSE\cite{pkg3}, and later
applied to generalized NLSE\cite{pkg4}.
The form of $M$ is restricted to be
\bea
M = \frac{{\vert \alpha \vert} {\vert \beta \vert} }{ \Gamma} \sigma_0
\sin(\theta_{\alpha}-\theta_{\beta}) + \alpha^{*} \sigma_{+} +
\alpha \sigma_{-}, \ \alpha={\vert \alpha \vert} e^{i \theta_{\alpha}},
\beta={\vert \beta \vert} e^{i \theta_{\beta}}
\eea
\noindent in order $A(t) = \mu_0(t) [ \beta^{*} \sigma_{+} + \beta \sigma_{-}
+ i \Gamma \sigma_3 ] $ to be $M$-pseudo-hermitian\cite{pkg3}. 
The eigenvalues of the matrix $M$ are $\lambda_{\pm}= {\vert \alpha \vert}
\left [ \frac{\vert \beta \vert}{\Gamma} \sin(\theta_{\alpha}-\theta_{\beta})
\pm 1 \right ]$ are real and can be positive as well as negative depending on
the values of the parameters.
We can distinguish three different region for $\lambda_{\pm}$. \\
{\bf{\underline{Region-I ( $\lambda_{\pm} > 0$) }}:} \\
Both the eigen values are positive when $\sin(\theta_\alpha - \theta_\beta) > \frac{\Gamma}{\vert \beta \vert}$, i.e. 
$\vert \beta \vert$ must be greater than $\vert \Gamma \vert$. In this region, 
for $\Gamma > 0$, $\sin(\theta_\alpha - \theta_\beta) > 0$ and for 
$\Gamma < 0$, $\sin(\theta_\alpha - \theta_\beta) < 0$. \\
{\bf{\underline{Region-II ($ \lambda_{+} > 0 \ \& \ \lambda_{-} < 0$)}}:} \\
When  the value of $\sin(\theta_\alpha - \theta_\beta)$ lies between $\frac{-\Gamma}{\vert \beta \vert}$ 
and $\frac{\Gamma}{\vert \beta \vert}$ i.e. $-\frac{\Gamma}{\vert \beta \vert} < \sin(\theta_{\alpha}-\theta_{\beta}) 
< \frac{\Gamma}{\vert \beta \vert}$, then $ \lambda_{+} > 0$ and $\lambda_{-} < 0$. Unlike region-I and III, 
$\vert \beta \vert$ may or may not be greater than $\vert \Gamma \vert$.  \\
{\bf{\underline{Region-III ($ \lambda_{\pm} < 0$)}}:} \\
Both the eigen values are negative when $\sin(\theta_\alpha - \theta_\beta) 
< \frac{-\Gamma}{\vert \beta \vert}$ i.e. $\vert \beta \vert > \vert \Gamma \vert$. For $\Gamma > 0$ , 
$\sin(\theta_\alpha - \theta_\beta) < 0$ and for $\Gamma < 0$, 
$\sin(\theta_\alpha - \theta_\beta) > 0$. \\
It is not possible to have $\lambda_{+} < 0$ and $\lambda_{-} > 0$.  
We are not restricting $M$ to be positive-definite, since
there is no requirement of $M$ to be interpreted as a metric on the associated
eigen-vector space of $A$ as is the case with quantum mechanics involving
pseudo-hermitian operator. Further, $M$ is not the unique matrix for showing $A$ to be
pseudo-hermitian, since the complex parameter $\alpha$ does not appear in the
matrix $A$ and can be chosen independently. Thus, the space of eigen vectors
associated with $A$ may be endowed with a positive-definite metric, 
\bea
\eta_+= \frac{{\vert \tilde{\alpha} \vert} {\vert \beta \vert} }{ \Gamma} \sigma_0
\sin(\theta_{\tilde{\alpha}}-\theta_{\beta}) + \tilde{\alpha}^{*} \sigma_{+} +
\tilde{\alpha} \sigma_{-},
\label{metric2}
\eea
\noindent  where $\alpha$ and $\tilde{\alpha}={\vert \tilde{\alpha} \vert} 
e^{i \theta_{\tilde{\alpha}}}$ are independent parameters. The positivity
of $\eta_+$ is ensured by choosing ${\vert \beta \vert} > {\vert \Gamma \vert}$
and $\theta_{\tilde{\alpha}}=\theta_{\beta} + (2n+1) \frac{\pi}{2}$ with even
$n$ for $\Gamma > 0$ and odd $n$ for $\Gamma < 0$. The unitary matrix,
\bea
\Bar{S} = \frac{1}{\sqrt{2}} \left [ \sigma_0 - e^{-i \theta_{\alpha}} \sigma_{+}
+ e^{i \theta_{\alpha}} \sigma_{-} \right ]
\label{unitary}
\eea
\noindent diagonalizes $M$, i.e. $\Bar{S}^{\dagger} M \Bar{S} =
\textrm{diag}(\lambda_+, \lambda_-)$. We express Eq. (\ref{p-herm2}) in terms of a two-component complex scalar
field $Q(x,t)$ by defining  $\Phi(x,t) = \Bar{S} \ Q(x,t)$, 
\bea
i Q_{t}(x,t) & = & - Q_{xx}(x,t) + V(x,t) Q(x,t) +
\left [\lambda_+ Q_1^{*}(-x,t)Q_1(x,t) \right . \nonumber \\
& + & \left .  \lambda_- \ Q_2^{*}(-x,t)Q_2(x,t) \right ] \ Q(x,t)
\label{p-herm3}
\eea
\noindent  Several solutions of Eq. (\ref{p-herm3}) with $V(x,t) = 0$ are
discussed in Sec. 3 of Ref. \cite{Khare} in terms Lam$\acute{e}$ Polynomials
of order 1 and 2. The exact solutions of Eq. (\ref{se1}) with $V(x)=0$ and
$A=A_0$ can be constructed corresponding to many of these solutions
by using the relation $\Psi=U \Bar{S} Q$ and Eqs. (\ref{punitary},
\ref{unitary}). It should be emphasized here that each and every solutions of Ref. \cite{Khare} can
not be mapped to be exact solutions of Eq. (\ref{p-herm3}), since the parameters $\lambda_{\pm}$
appearing in the latter equation have specific forms and do not cover the whole of the parameter-space.
$\lambda_+ < 0$ and $\lambda_->0$  are forbidden for Eq. (\ref{p-herm3}), while
the corresponding equation of Ref. (\cite{Khare}) admits exact analytical solutions in the respective
parameter-space. For example, There is no solution of Eq. (\ref{p-herm3}) with $V(x,t)=0$ 
corresponding to the Solution IV and V in Sec-3 of  Ref. \cite{Khare}. We discuss two independent 
solutions in terms of hyperbolic functions which are
obtained as limiting cases of Lam$\acute{e}$ Polynomials.
In the Appendix-I, several exact solutions for arbitrary $\mu(t)$ are presented. 

The explicit expression of the solution of Eq. (\ref{nlse1}) is,
\bea
\psi_1(x,t) & = & \frac{1}{\sqrt{2}} \ \big[ \{ cos(\epsilon) +
(\Gamma - i\beta^{*} e^{i\theta_{\alpha}}) \frac{\sin(\epsilon)}{\epsilon_0} \}
Q_1(x,t) \nonumber \\
            & + & \{-e^{-i\theta_{\alpha}} \cos(\epsilon) -
(\Gamma e^{-i\theta_{\alpha}} + i \beta^{*}) \frac{\sin(\epsilon)}{\epsilon_0} \}
Q_2  \big] \nonumber \\
\psi_2(x,t) & = & \frac{1}{\sqrt{2}} \ \big[ \{ \cos(\epsilon) -
(\Gamma - i\beta e^{-i\theta_{\alpha}}) \frac{\sin(\epsilon)}{\epsilon_0} \}
Q_2(x,t) \nonumber \\
            & + & \{e^{i\theta_{\alpha}} \cos(\epsilon) - (\Gamma e^{i\theta_{\alpha}}
+ i \beta) \frac{\sin(\epsilon)}{\epsilon_0} \} Q_1(x,t) \big] 
\eea
\noindent This is the general expression of $\psi_{j}$ for any value of $\epsilon_0$. In the limit 
$\epsilon_0 \rightarrow 0$, $\cos(\epsilon) \rightarrow 1$ and 
$\frac{\sin(\epsilon)}{\epsilon_0} \rightarrow \mu(t)$. We can get the expression of $\psi_{j}$ for purely 
imaginary $\epsilon_0$, replacing $\epsilon_0$ by $i|\epsilon_0|$. 
The appropriate choice of $\mu_0(t)$ ensure the bounded nature of the solution of Eq. (\ref{nlse1}) in 
the limit $\epsilon_0$ is zero and purely imaginary. A large no of $\mu_0(t)$ is possible for different 
regions of $\epsilon_0$. This BLG-LC method can be used to stabilize the solutions of nonlocal nonlinear 
system as discussed in Ref. \cite{winfal}. 

\subsubsection{$V(x,t)=0$: Vanishing confining potential}
In this subsection we present a few exact solutions with specific choice of $\mu_0$.
In the Appendix-I, several exact solutions for arbitrary $\mu(t)$ are presented. 
\\
\noindent{\bf Example-I}: We present bright-bright soliton solution of Eq. (\ref{p-herm3})
with $V(x,t)=0$ corresponding to the Solution XXII in Ref. \cite{Khare}
\bea
&& Q_1(x,t) = [A \sech^2(\Omega x) + D] e^{-i(\omega_1 t + \delta)}, \ Q_2(x,t) =
B \sech^2(\Omega x) e^{-i(\omega_2 t + \delta)} \nonumber \\
&& \lambda_{-} B^2 = -\lambda_{+} A^2 = -\frac{9}{2} \Omega^2 \nonumber \\
&& \frac{D}{A} = - \frac{2}{3}, \ \omega_1 = - \omega_2 = 2 \Omega^2
\eea
\noindent where $A$ and $B$ denote the amplitudes and $\Omega^2$ is a positive.
Hence the solution is applicable only when $\lambda_{+} > 0$ and $\lambda_{-}
< 0$ i.e. $-\frac{\Gamma}{\vert \beta \vert} < \sin(\theta_{\alpha} - \theta_{\beta})
<  \frac{\Gamma}{\vert \beta \vert}$ condition is satisfied. The solution of Eq. (\ref{nlse1}) is,
\bea
\psi_1(x,t) & = & \frac{1}{\sqrt{2}} \ \left[ \left\{ \cos(\epsilon) + (\Gamma - i\beta^{*} e^{i\theta_{\alpha}})
\frac{\sin(\epsilon)}{\epsilon_0} \right\} \ \left\{A \sech^2(\Omega x) + D\right\} \ e^{-i(\omega_1 t + \delta)} \right. \nonumber \\
	    & + &\left. \left\{-e^{-i\theta_{\alpha}} \cos(\epsilon)  - (\Gamma e^{-i\theta_{\alpha}} + i \beta^{*})
\frac{\sin(\epsilon)}{\epsilon_0} \right\} \ B \sech^2(\Omega x) \ e^{-i(\omega_2 t + \delta)} \right] \nonumber \\
\psi_2(x,t) & = & \frac{1}{\sqrt{2}} \ \left[ \left\{ \cos(\epsilon) - (\Gamma - i\beta e^{-i\theta_{\alpha}})
\frac{\sin(\epsilon)}{\epsilon_0} \right\} \ B \sech^2(\Omega x) \ e^{-i(\omega_2 t + \delta)} \right. \nonumber \\
	    & + & \left. \left\{e^{i\theta_{\alpha}} \cos(\epsilon) - (\Gamma e^{i\theta_{\alpha}} + i \beta)
\frac{\sin(\epsilon)}{\epsilon_0} \right\} \ \left\{A \sech^2(\Omega x) + D \right\} \ e^{-i(\omega_1 t + \delta)} \right] \nonumber \\
\eea

\begin{figure}[!h]
	\begin{subfigure}{0.32\textwidth}
		\centering
		\includegraphics[width=0.99\linewidth]{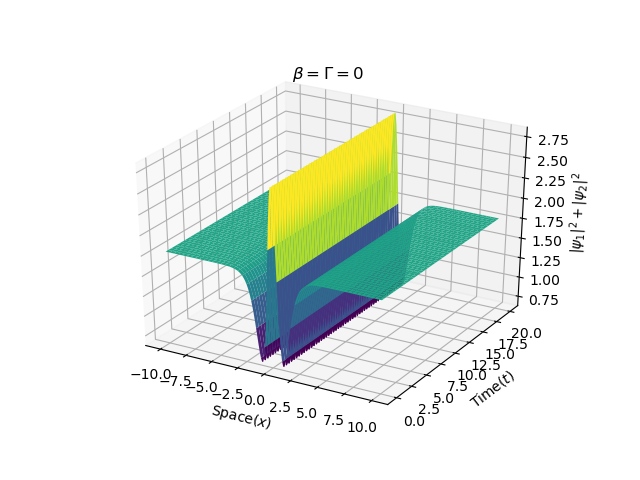}\quad
		\caption{}
		\label{fig:1}
	\end{subfigure}
	\begin{subfigure}{0.32\textwidth}
                \centering
                \includegraphics[width=0.99\linewidth]{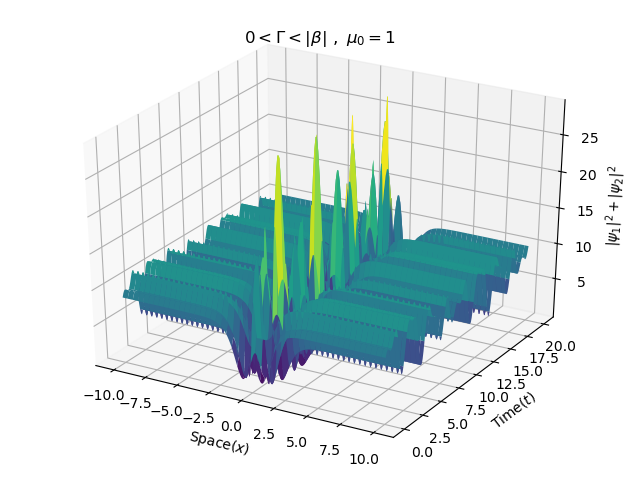}\quad
                \caption{}
                \label{fig:2}
        \end{subfigure}
	\begin{subfigure}{0.32\textwidth}
                \centering
                \includegraphics[width=0.99\linewidth]{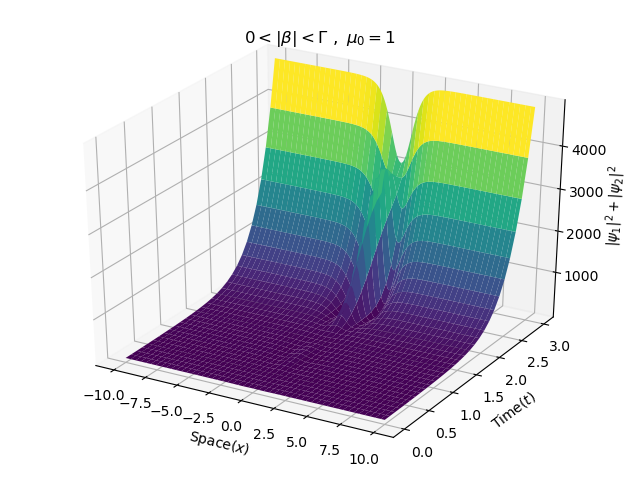}\quad
                \caption{}
                \label{fig:3}
        \end{subfigure}
	\medskip
	\begin{subfigure}{0.32\textwidth}
                \centering
                \includegraphics[width=0.99\linewidth]{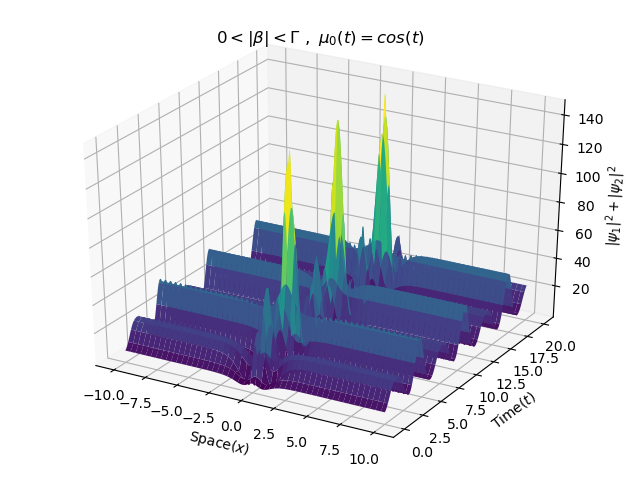}\quad
                \caption{}
                \label{fig:4}
        \end{subfigure}
	\begin{subfigure}{0.32\textwidth}
                \centering
                \includegraphics[width=0.99\linewidth]{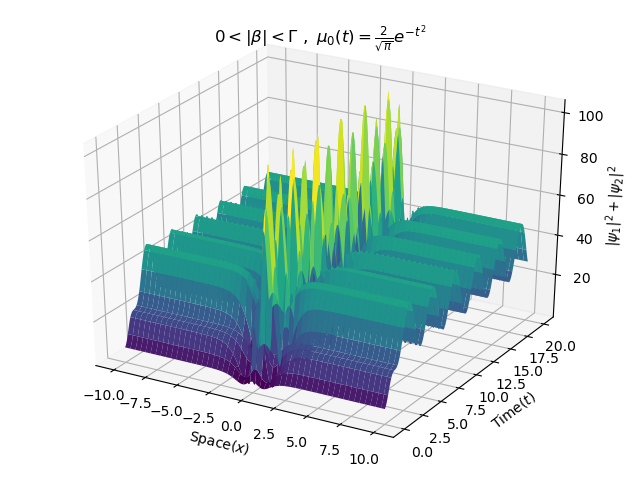}\quad
                \caption{}
                \label{fig:5}
        \end{subfigure}
	\begin{subfigure}{0.32\textwidth}
                \centering
                \includegraphics[width=0.99\linewidth]{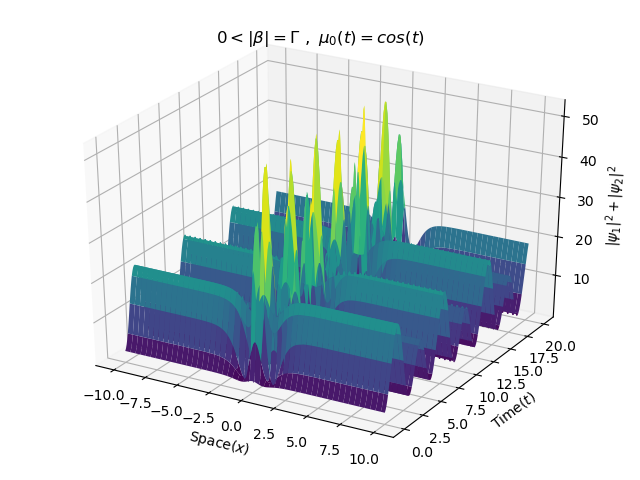}\quad
                \caption{}
                \label{fig:6}
        \end{subfigure}
	\caption{ (Color online) Plot of $|\psi_1|^{2} + |\psi_2|^{2}$ of example-I  
	for the parametric value $\theta_{\beta} = \frac{\pi}{3}$, 
	$\theta_{\alpha} =\frac{\pi}{6}$, $\omega_1 = 2$, $\omega_2 = -2$, $\Omega = 1$ 
	and $\delta = 0$. In Fig(a): $\Gamma = |\beta| =0, A = 2.121, B = 2.121$, 
	Fig(b): $\Gamma = 1.5, |\beta| = 2, \mu_0 = 1, A = 3.674, B = 1.643 $, Fig(c): $\Gamma =2.3, |\beta|= 2,
	\mu_0 =1, A = 2.822, B = 1.771$, Fig(d): $\Gamma=2.3, |\beta|=2,\mu_0(t) = cos(t), A = 2.822, 
	B = 1.771$, Fig(e): $\Gamma=2.3, |\beta|=2, \mu_0(t) = \frac{2}{\sqrt{\pi}} e^{-t^2}, 
	A = 2.822, B = 1.771$, Fig(f): $\Gamma=|\beta|=2,\mu_0(t) = cos(t), A = 3, B = 1.732$}
	\label{img1}
\end{figure}

Figure. 1 shows the power $P = |\psi_1|^2 + |\psi_2|^2$ plotted, for various time-modulations. 
Fig. 1(a) depicts the scenario when BLG and LC vanish, and there is no change in the power over time.
For constatnt $\mu_0$ and a non-vanishing $\beta$,$\Gamma$, fulfilling $|\beta| > \Gamma$, the 
power-oscillation can be shown in Fig. 1(b). The power oscillation in time is the signature of the 
balanced loss-gain system. The solution $\Psi$ becomes unstable for constant $\mu_0$ and non-vanishing 
$\beta$, \ $\Gamma$ while satisfying $|\beta| \leq \Gamma$. The relevant plot is displayed in Fig. 1.(c). 
It may be recalled from our earlier discussions in Sec. 3 that the transformation
(\ref{unitary1}) that joins the initial system given by Eq. (\ref{nlse2}) to Eq. (\ref{nlse3}),
does not change the stability property of $\Phi(x,t)$ for a bounded U(t). Thus, for a given stable solution 
of Eq. (\ref{nlse3}), the solutions of Eq. (\ref{nlse2}) are stable for bounded $U(t)$, i.e. suitable choice of $\mu_0(t)$. 
We can control the instability by wisely selecting $\mu_0(t)$. 
We have taken into account $|\beta| < \Gamma$ and a periodic modulation function $\mu_0(t) = cos(t)$ 
for which $P(x,t)$ exhibits periodic behaviour in Fig. 1(d). In Fig. 1(e), which is likewise time-bounded, 
we have plotted power for the conditions $|\beta| < \Gamma$ and $\mu_0(t) = \frac{2}{\sqrt{\pi}} e^{-t^2}$. 
The instability in the region $|\beta| = \Gamma$ may once more be controlled using periodic time 
moduation. The plot in Fig. 1(f) in particular displays power oscillation for $\mu_0(t) = cos(t)$ in the 
region $|\beta| = \Gamma$ . The mentioned features have been verified up to time $t \approx 200$, 
although the figures are only displayed up to time $t \approx 20$.

\noindent{\bf Example-II}: We present the bright-dark soliton solution of Eq. (\ref{p-herm3})
with $V(x)=0$ corresponding to the Solution VIII in Ref. \cite{Khare},
\bea
&& Q_1(x,t) = A \sech \left (\Omega x \right ) e^{-i \left (\omega_1 t + \delta
\right )}, \ \
Q_2(x,t) = B \tanh \left (\Omega x \right ) e^{-i \left (\omega_2 t + \delta
\right )},\nonumber \\
&& \Omega^2 = -\frac{1}{2}(\lambda_{+} A^2 + \lambda_{-} B^2),\nonumber \\
&& \omega_1=\Omega^2 + \lambda_+ A^2, \ \ \omega_2=\Omega^2+\omega_1
\eea
\noindent where $A$ and $B$ denote the amplitudes. The positivity of $\Omega^2$
is ensured when $\frac{B^2-A^2}{B^2+A^2} > \frac{|\beta|}{\Gamma} \ \sin(\theta_\alpha - \theta_\beta)$. 
The solution of Eq. (\ref{nlse1}) is,
\bea
\psi_1(x,t) & = & \frac{1}{\sqrt{2}} \ \left[ \left\{ cos(\epsilon) + 
(\Gamma - i\beta^{*} e^{i\theta_{\alpha}}) \frac{\sin(\epsilon)}{\epsilon_0} \right\}  
             A \ \sech(\Omega x) \ e^{-i(\omega_1 t + \delta)} \right. \nonumber \\ 
	    & + & \left. \left\{-e^{-i\theta_{\alpha}} \cos(\epsilon) - 
(\Gamma e^{-i\theta_{\alpha}} + i \beta^{*}) \frac{\sin(\epsilon)}{\epsilon_0} \right\} 
             \ B \ \tanh(\Omega x) \ e^{-i(\omega_2 t + \delta)} \right] \nonumber \\
\psi_2(x,t) & = & \frac{1}{\sqrt{2}} \ \left[ \left\{ cos(\epsilon) - 
(\Gamma - i\beta e^{-i\theta_{\alpha}}) \frac{\sin(\epsilon)}{\epsilon_0} \right\} 
             B \ \tanh(\Omega x) \ e^{-i(\omega_2 t + \delta)} \right. \nonumber \\ 
	    & + & \left. \left\{e^{i\theta_{\alpha}} \cos(\epsilon) - (\Gamma e^{i\theta_{\alpha}} 
+ i \beta) \frac{\sin(\epsilon)}{\epsilon_0} \right\} 
             A \ \sech(\Omega x) \ e^{-i(\omega_1 t + \delta)} \right] 
\eea

\begin{figure}[!h]
        \begin{subfigure}{0.32\textwidth}
                \centering
                \includegraphics[width=0.99\linewidth]{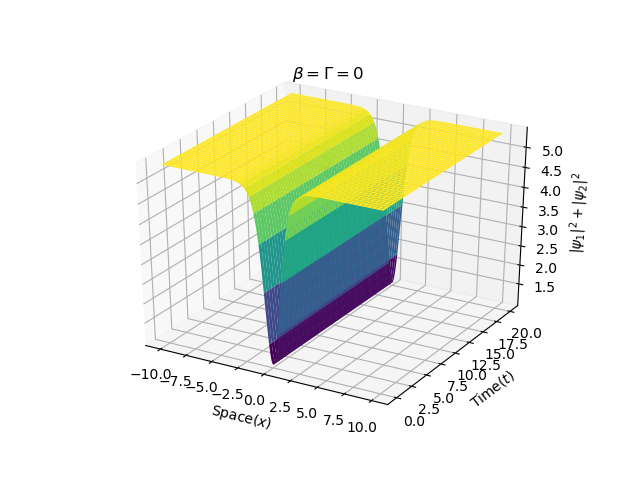}\quad
                \caption{}
                \label{fig:7}
        \end{subfigure}
        \begin{subfigure}{0.32\textwidth}
                \centering
                \includegraphics[width=0.99\linewidth]{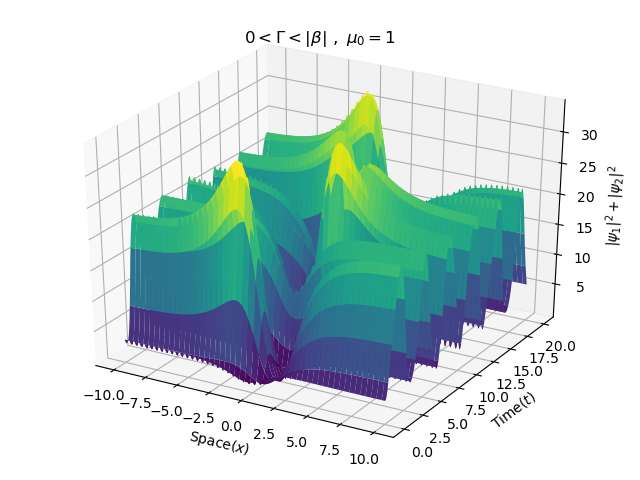}\quad
                \caption{}
                \label{fig:8}
        \end{subfigure}
        \begin{subfigure}{0.32\textwidth}
                \centering
                \includegraphics[width=0.99\linewidth]{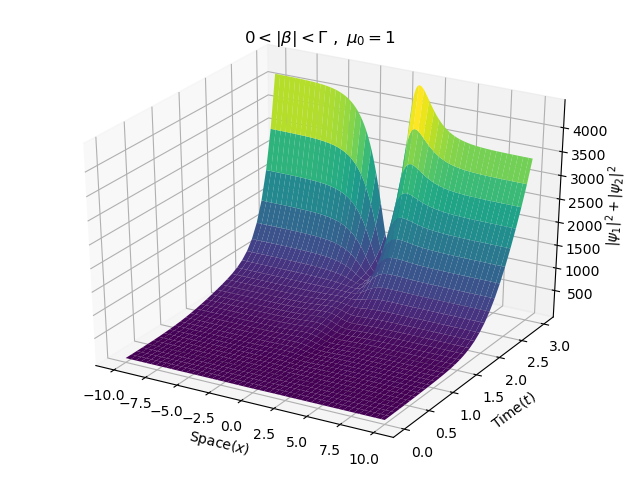}\quad
                \caption{}
                \label{fig:9}
        \end{subfigure}
        \medskip
        \begin{subfigure}{0.32\textwidth}
                \centering
                \includegraphics[width=0.99\linewidth]{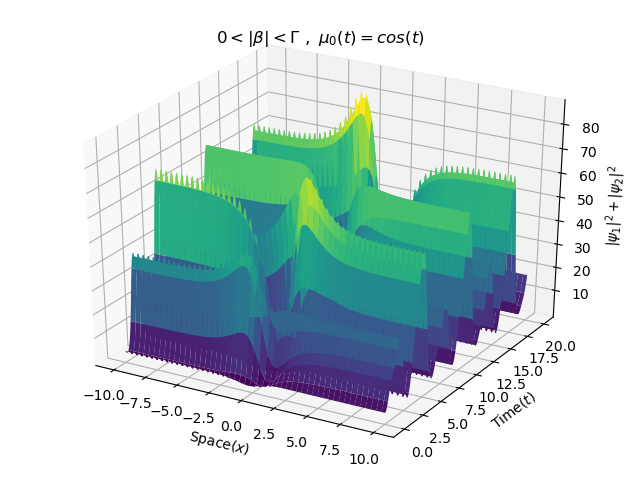}\quad
                \caption{}
                \label{fig:10}
        \end{subfigure}
        \begin{subfigure}{0.32\textwidth}
                \centering
                \includegraphics[width=0.99\linewidth]{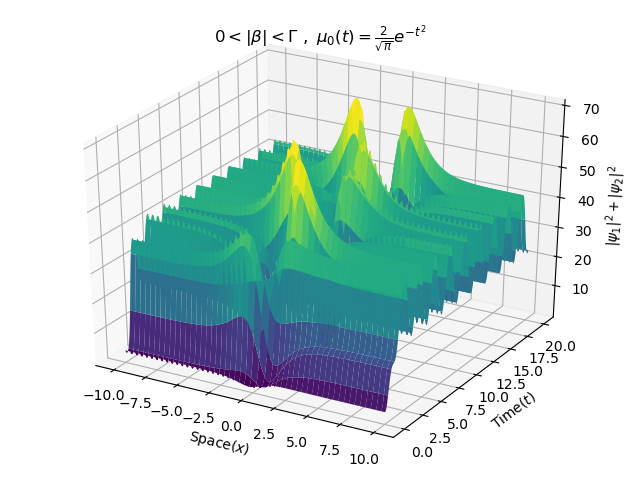}\quad
                \caption{}
                \label{fig:11}
        \end{subfigure}
        \begin{subfigure}{0.32\textwidth}
                \centering
                \includegraphics[width=0.99\linewidth]{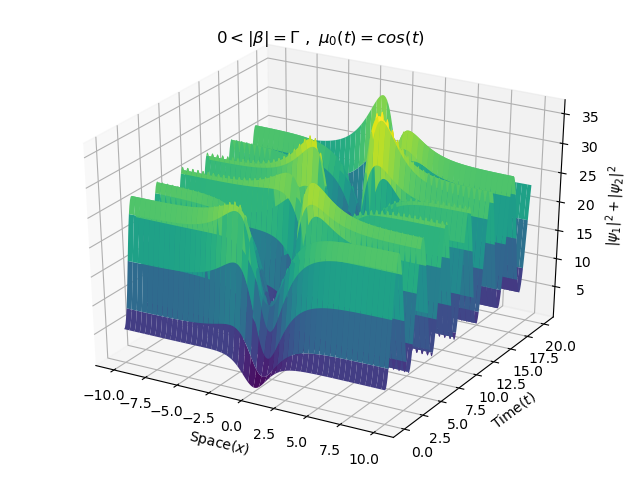}\quad
                \caption{}
                \label{fig:12}
        \end{subfigure}
        \caption{ (Color online) Plot of $|\psi_1|^{2} + |\psi_2|^{2}$ of example-II
	for the parametric value $A=1$, $B =\sqrt{\frac{27}{5}}$, $\theta_{\beta} = \frac{\pi}{6}$,
        $\theta_{\alpha} =\frac{\pi}{3}$ and $\delta = 0$. In Fig(a): $\Gamma = |\beta| =0, \omega_1 = 3.2, 
	\omega_2 = 5.4, \Omega = 2.2$, Fig(b): $\Gamma = 1.5, |\beta| = 2, \mu_0 = 1, \omega_1 = 1.957, 
	\omega_2 = 2.349, \Omega = 0.626$, Fig(c): $\Gamma =2.3, |\beta|= 2,
        \mu_0 =1, \omega_1 = 2.243, \omega_2 = 3.052, \Omega = 0.899$, Fig(d): $\Gamma=2.3, |\beta|=2,\mu_0(t) = cos(t), 
	\omega_1 = 2.243, \omega_2 = 3.052, \Omega = 0.899$, Fig(e): $\Gamma=2.3, |\beta|=2, \mu_0(t)
        = \frac{2}{\sqrt{\pi}} e^{-t^2}, \omega_1 = 2.243, \omega_2 = 3.052, \Omega = 0.899$, 
	Fig(f) : $\Gamma=|\beta|=2,\mu_0(t) = cos(t), \omega_1 = 2.1, \omega_2 = 2.7, \Omega = 0.774$}
        \label{img2}
\end{figure}

The qualitative behaviour of the plots in Figure (2) under the same circumstances on $\beta, \Gamma$ and $\mu_0(t)$ 
is identical to that of Fig. 1. The power oscillation is absent for vanishing $\beta,\Gamma$ as shown in 
Fig. 2(a). Fig. 2(b) depicts the power-oscillation for constant $\mu_0$ and $|\beta| > \Gamma$. As seen in Fig. 2(c), 
the solution expands with no upper bound for constants $\mu_0$ and $|\beta| < \Gamma$. Figs. 2(d), 2(e), and 2(f) 
illustrate how instabilities for $|\beta| \leq \Gamma$ are managed by selecting an appropriate $\mu_0(t)$.

\subsubsection{Non-vanishing confining potential}

We have presented two solutions of Eq. (\ref{p-herm3}) for a special case $V(x,t) = 0$. Several solutions of 
Eq. (\ref{p-herm3}) with $\cal{PT}$-symmetric confining potential can be constructed. The solution 
of Eq. (\ref{p-herm3}) with $V(x,t) = S(x) + i \tilde{S}(x)$ can be written as,
\bea
Q_1(x,t) & = & W_1 \ q(x) \ e^{i\big(r(x) + \omega t + \theta_1 \big)} \nonumber \\
Q_2(x,t) & = & W_2 \ q(x) \ e^{i\big(r(x) + \omega t + \theta_2 \big)} 
\label{q-sol1}
\eea
\noindent where $q(x)$ and $r(x)$ are real function and denote the amplitude and phase respectively.  
We substitute Eq. (\ref{q-sol1}) into Eq. (\ref{p-herm3}) and the  amplitude and phase satisfy the relations 
\bea
S(x)         & = & -\omega  + \frac{q_{xx}}{q} - r^{2}_{x} - \alpha \ q(x) \ q(-x) \ \cos(y(x)) \nonumber \\
\tilde{S}(x) & = & r_{xx} + \frac{2 q_{x} r_{x}}{q} - \alpha \ q(x) \ q(-x) \ \sin(y(x)) 
\label{q-sol2}
\eea
\noindent where $y(x) = r(x) - r(-x)$ and the constant $\alpha = \lambda_{+} |W_1|^{2} + \lambda_{-} |W_2|^{2}$. 
The solution of Eq. (\ref{q-sol2}) for different kinds of $\cal{PT}$-symmetric potential is discussed in 
Ref. \cite{Wen}.  The exact solutions of Eq. (\ref{se1}) with $V(x,t) = S(x) + i \tilde{S}(x)$ can be 
constructed corresponding to each of these solutions by using the relation 
$\Psi = U \Bar{S} W q(x) e^{i\big(r(x)+ \omega t\big)}$ where the column matrix 
$W = \Big( W_1 e^{i\theta_1} \ \ W_2 e^{i\theta_2} \Big)^{T}$. 

\noindent{\bf Generalised Rosen-Morse Potential}: We first consider $\cal{PT}$-symmetric generalised
complex Rosen-Morse Potential $S(x)+ i\tilde{S}(x)$ with the components 
\bea
S(x)         & = & - 2 \sech^{2}(x) - \alpha a^2 \sech^{2}(x) \cos(y(x)) \nonumber \\
\tilde{S}(x) & = & - 2b \tanh(x) - \alpha a^2 \sech^{2}(x) \sin(y(x))
\eea
\noindent where $y(x) = 2bx$ and $a,b$ are constant. The solution of Eq. (\ref{p-herm3}) corresponding to this 
potential is 
\bea
Q(x,t) = \begin{bmatrix}
	W_1 e^{i\theta_1} \\
	W_2 e^{i\theta_2}
\end{bmatrix} \ a \sech(x) \ e^{i\big(bx + (1-b^2)t \big)}
\eea
\noindent We can choose appropriate physically motivated  $\mu_0(t)$ so that $U(t)$ is bounded. We present 
the solution of Eq. (\ref{nlse1}) for this generalised Rosen-Morse potential and for a given $\mu_0(t)$.
\bea
\psi_1(x,t) & = & \frac{1}{\sqrt{2}} \ \big[ \{ cos(\epsilon) +
(\Gamma - i\beta^{*} e^{i\theta_{\alpha}}) \frac{\sin(\epsilon)}{\epsilon_0} \} 
W_1 \ e^{i\theta_1} \nonumber \\
            & + & \{-e^{-i\theta_{\alpha}} \cos(\epsilon) -
(\Gamma e^{-i\theta_{\alpha}} + i \beta^{*}) \frac{\sin(\epsilon)}{\epsilon_0} \} 
W_2 \ e^{i\theta_2}  \big] \nonumber \\
            &   & \times \ a \sech(x) \ e^{i\big(bx + (1-b^2)t \big)}  \nonumber \\
\psi_2(x,t) & = & \frac{1}{\sqrt{2}} \ \big[ \{ \cos(\epsilon) -
(\Gamma - i\beta e^{-i\theta_{\alpha}}) \frac{\sin(\epsilon)}{\epsilon_0} \} 
W_2 \ e^{i \theta_2} \nonumber \\
            & + & \{e^{i\theta_{\alpha}} \cos(\epsilon) - (\Gamma e^{i\theta_{\alpha}}
+ i \beta) \frac{\sin(\epsilon)}{\epsilon_0} \} W_1 \ e^{i\theta_1} \big] \nonumber \\
            &   & \times \ a \sech(x) \ e^{i\big(bx + (1-b^2)t \big)}
\eea

\noindent{\bf Generalised Scarf-II potential}: We consider $\cal{PT}$-symmetric generalised
Scarf-II potential. The real and imaginary parts of the potential are, 
\bea
S(x)         & = & - (2 + b^2) \sech^2(x) - \alpha a^2 \sech^2(x) \cos(y(x)) \nonumber \\
\tilde{S}(x) & = & - 3b \sech(x) \tanh(x) - \alpha a^2 \sech^2(x) \sin(y(x))
\eea
\noindent where $y(x) = 2b \arctan[\sinh(x)]$ and $a,b$ are constant. The solution of 
Eq. (\ref{p-herm3}) corresponding to this potential
\bea
Q(x,t) = \begin{bmatrix}
        W_1 e^{i\theta_1} \\
        W_2 e^{i\theta_2}
\end{bmatrix} \ a \sech(x) \ e^{i\big(b \arctan[\sinh(x)] + t \big)}
\eea
\noindent We present the solution of Eq. (\ref{nlse1}) for this generalised Scarf-II potential 
and for a given $\mu_0(t)$.
\bea
\psi_1(x,t) & = & \frac{1}{\sqrt{2}} \ \big[ \{ cos(\epsilon) +
(\Gamma - i\beta^{*} e^{i\theta_{\alpha}}) \frac{\sin(\epsilon)}{\epsilon_0} \}
W_1 \ e^{i\theta_1} \nonumber \\
            & + & \{-e^{-i\theta_{\alpha}} \cos(\epsilon) -
(\Gamma e^{-i\theta_{\alpha}} + i \beta^{*}) \frac{\sin(\epsilon)}{\epsilon_0} \}
W_2 \ e^{i\theta_2}  \big] \nonumber \\
            &   & \times \ a \sech(x) \ e^{i\big(b \arctan[\sinh(x)] + t \big)}  \nonumber \\
\psi_2(x,t) & = & \frac{1}{\sqrt{2}} \ \big[ \{ \cos(\epsilon) -
(\Gamma - i\beta e^{-i\theta_{\alpha}}) \frac{\sin(\epsilon)}{\epsilon_0} \}
W_2 \ e^{i \theta_2} \nonumber \\
            & + & \{e^{i\theta_{\alpha}} \cos(\epsilon) - (\Gamma e^{i\theta_{\alpha}}
+ i \beta) \frac{\sin(\epsilon)}{\epsilon_0} \} W_1 \ e^{i\theta_1} \big] \nonumber \\
            &   & \times \ a \sech(x) \ e^{i\big(b \arctan[\sinh(x)] + t \big)}
\eea

\noindent{\bf Periodic Potential}: We consider $\cal{PT}$-symmetric generalised periodic
potential. The real and imaginary parts of the potential are,
\bea
S(x)         & = & - b^2 \cos^2(x) - \alpha a^2 \cos^2(x) \cos(y(x)) \nonumber \\
\tilde{S}(x) & = & - 3b \sin(x) - \alpha a^2 \cos^2(x) \sin(y(x))
\eea
\noindent where $y(x) = 2b \sin(x)$ and $a,b$ are constant. The solution of
Eq. (\ref{p-herm3}) corresponding to this potential
\bea
Q(x,t) = \begin{bmatrix}
        W_1 e^{i\theta_1} \\
        W_2 e^{i\theta_2} 
\end{bmatrix} \ a \cos(x) \ e^{i\big(b \sin(x) + t \big)}
\eea
\noindent We present the solution of Eq. (\ref{nlse1}) for this $\cal{PT}$-symmetric periodic 
potential and for a given $\mu_0(t)$.
\bea
\psi_1(x,t) & =      & \frac{1}{\sqrt{2}} \ \big[ \{ cos(\epsilon) +
(\Gamma - i\beta^{*} e^{i\theta_{\alpha}}) \frac{\sin(\epsilon)}{\epsilon_0} \}
W_1 \ e^{i\theta_1} \nonumber \\
            & +      & \{-e^{-i\theta_{\alpha}} \cos(\epsilon) -
(\Gamma e^{-i\theta_{\alpha}} + i \beta^{*}) \frac{\sin(\epsilon)}{\epsilon_0} \}
W_2 \ e^{i\theta_2}  \big] \nonumber \\
	    & \times & \ a \cos(x) \ e^{i\big(b \sin(x) + t \big)}  \nonumber \\
\psi_2(x,t) & =      & \frac{1}{\sqrt{2}} \ \big[ \{ \cos(\epsilon) -
(\Gamma - i\beta e^{-i\theta_{\alpha}}) \frac{\sin(\epsilon)}{\epsilon_0} \}
W_2 \ e^{i \theta_2} \nonumber \\
            & +      & \{e^{i\theta_{\alpha}} \cos(\epsilon) - (\Gamma e^{i\theta_{\alpha}}
+ i \beta) \frac{\sin(\epsilon)}{\epsilon_0} \} W_1 \ e^{i\theta_1} \big] \nonumber \\
	    & \times & \ a \cos(x) \ e^{i\big(b \sin(x) + t \big)}
\eea

\subsection{General Case}

To solve the general case i.e $V(x,t) \neq 0$ and $M \neq L$, we consider the solution of Eq. (\ref{nlse3}) as,
\bea
\Phi(x,t) & = & W \rho(x,t) e^{i\eta(x,t)} u(\zeta) \ ; \ \zeta = \zeta(x,t) 
\label{ansatz1} 
\eea
where $W = \big( W_1 e^{i \theta_1} \ \ W_2 e^{i \theta_2} \big)^{T} $ is a constant complex vector 
and $\zeta(x,t)$ is an odd function under parity transformation i.e. $\zeta(-x,t) = -\zeta(x,t)$. 
This is an essential condition because our aim is to reduce Eq. (\ref{nlse3}) into a solvable equation and 
Eq. (\ref{nlse3}) reduces to solvable Eq. (\ref{u3}) only when $\zeta(-x,t)$ is an odd function under parity 
transformation. $\rho(x,t),\zeta(x,t),\eta(x,t)$ and $V(x,t)$ satisfy the following sets of equations
\bea
\rho\rho_{t} + (\rho^{2} \eta_{x})_{x} & = & \rho^{2}(x,t)\tilde{S}(x,t) \\
\label{con1}
\zeta_{t}(x,t) + 2 \eta_{x}(x,t) \zeta_{x}(x,t)       & = & 0  \\
\label{con2}
(\rho^{2} \zeta_{x})_{x}                              & = & 0  \\
\label{con3}
S(x,t)                                                & = & \frac{\rho_{xx}}{\rho} 
- \eta_{t} - \eta^{2}_{x} - \mu \zeta^{2}_{x} 
\label{con4}
\eea 
\noindent where $\mu$ is a constant. This set of equations is same as shown in Ref \cite{pkg2} in the context 
of one component NNLSE. The ansatz as shown in Eq. (\ref{ansatz1}), was also used in Ref. \cite{Beitia2} for 
one component localized NLSE. Inserting the Eq.( \ref{ansatz1}) into Eq. (\ref{nlse3}) we get,
\begin{eqnarray}
u_{\zeta\zeta} + \mu \ u  & = & \frac{1}{W^{\dagger}W} [ W^{\dagger} K(l_{11},m_{11})W W^{\dagger}F_{+}W  
+  W^{\dagger}K(l_{22},m_{22})W W^{\dagger}F_{-}W \nonumber \\ 
                      & + & W^{\dagger} K(l_{12},m_{12})W W^{\dagger}BW  
+  W^{\dagger}K(l_{21},m_{21})W W^{\dagger}B^{\dagger}W] \nonumber \\
                      & \times & \frac{\rho(-x,t) \rho(x,t)}{\zeta^{2}_{x}} \ 
e^{i(\eta(x,t)-\eta(-x,t))} \ u^{*}(-\zeta)u^{2}(\zeta)
\label{u1}
\end{eqnarray}
\noindent The nonlinear term will be real for the judicious choice of $l_{ij}$ and $m_{ij}$. To
ensure that the co-efficient of the nonlinear term is real and Eq. (\ref{u1}) reduces to 
Eq. (\ref{u2}), the following conditions should be satisfied
\bea
&& l_{12} = l_{21} \ ; \ m_{12} = m_{21} \nonumber \\
&& l_{11} + m_{11} = \frac{f_1(x,t)}{W^{\dagger}F_{+}W} \nonumber \\
&& l_{22} + m_{22} = \frac{f_2(x,t)}{W^{\dagger}F_{-}W} \nonumber \\
&& l_{12} + m_{12} = \frac{f_3(x,t)}{W^{\dagger}(B+B^{\dagger})W} \nonumber \\
&& l_{11} - m_{11} =  - (l_{22} - m_{12}) \frac{W^{\dagger}F_{-}W}{W^{\dagger}F_{+}W} 
-(l_{12} - m_{12}) \frac{ W^{\dagger}(B+B^{\dagger})W}{W^{\dagger}F_{+}W} 
\label{l-eqn1}
\eea
\noindent where $f_1(x,t)$,$f_2(x,t)$ and $f_3(x,t)$ are arbitrary functions. The expressions of 
space-time modulated strength $l_{ij}$,$m_{ij}$ of nonlinear interaction may be obtained by solving
the group of Eqs. (\ref{l-eqn1}).  We solve the equations by keeping $m_{22}$, $m_{12}$ and $m_{21}$ 
arbitrary :
\bea
l_{11} & = & \frac{1}{2W^{\dagger}F_{+}W} [ f_1(x,t) - f_{2}(x,t) - f_{3}(x,t) \nonumber \\
       & + & 2 m_{22} W^{\dagger}F_{-}W + 2 m_{12} W^{\dagger}(B+B^{\dagger})W] \nonumber \\
m_{11} & = & \frac{1}{2W^{\dagger}F_{+}W} [ f_1(x,t) + f_{2}(x,t) + f_{3}(x,t) \nonumber \\
       & - & 2 m_{22} W^{\dagger}F_{-}W - 2 m_{12} W^{\dagger}(B+B^{\dagger})W] \nonumber \\
l_{12} & = & \frac{f_3(x,t)}{W^{\dagger}(B+B^{\dagger})W} - m_{12} \nonumber \\
l_{22} & = & \frac{f_2(x,t)}{W^{\dagger}F_{-}W} - m_{22} 
\label{l-eqn2}
\eea
\noindent The above expressions will be same for both the local and nonlocal cases. For 
$l_{ij} = m_{ij} = 0$ , $i \neq j$ the Eq. (\ref{l-eqn2}) reduces to the Eq. (19) of Ref. \cite{pkg4}.
The signature of the nonlocality in Eq. (\ref{u1}) is carried by the terms $\rho(-x,t)$, $\eta(-x,t)$ and $u^{*}(-\zeta)$. 
Instead of the complicated form we choose a simple form of $l_{ij}$, $m_{ij}$ to present our result: 
\begin{eqnarray}
l_{11} = \frac{f_1(x,t)+G(x,t)}{2 W^{\dagger}F_{+}W} \ ; 
\ m_{11} = \frac{f_1(x,t) - G(x,t)}{2 W^{\dagger}F_{+}W} \nonumber \\
l_{22} = \frac{f_{2}(x,t)-G(x,t)}{2 W^{\dagger}F_{-}W} \ ; 
\ m_{22} = \frac{f_{2}(x,t)+G(x,t)}{2 W^{\dagger}F_{-}W} \nonumber \\
l_{21} = l_{12} = m_{12} = m_{21} = \frac{f_3(x,t)}{2W^{\dagger}(B+B^{\dagger})W}
\label{g4}
\end{eqnarray}
\noindent where $G(x,t)$ is an arbitrary function. 
The choice of $l_{ij}$ and $m_{ij}$ to ensure that the nonlinear term is real, is not unique. 
There are others choices too. For the time being we are considering the above choices.
The expressions for $W^{\dagger}F_{\pm}W$ are 
\begin{eqnarray}
W^{\dagger} F_{\pm} W = b_0 \big(\frac{1}{2} + T_{\pm} \mu(t) D \big) + b_{3} 
[\frac{\Gamma \mu(t)}{2 \epsilon} \sin(2 \epsilon) \pm \frac{1}{2} \cos(2 \epsilon)]
\end{eqnarray}
where $\beta = \vert \beta \vert e^{i \theta_3}$ and $b_{j} = W^{\dagger} \sigma_{j} W \ ; 
\ j = 0,1,2,3$. The constant $D$ is defined as,
\begin{eqnarray}
D = \Gamma + \frac{2}{b_0} \ W_1 \ W_2 \vert \beta \vert \sin(\theta_2 - \theta_1 -\theta_3) 
\end{eqnarray}
Both the expressions of $W^{\dagger} F_{\pm} W$ and $D$ are shown in the Ref.\cite{pkg4} in the  
context of local NLSE. In both the cases the above expressions are same. Due to the presence of 
cross terms $l_{ij},m_{ij}$ with $i \neq j$ the terms $W^{\dagger} B W$ and $W^{\dagger} B^{\dagger} W$ 
have arisen. For the nonlinear term to be real, as we have considered all the cross terms 
are equal to each other, from the last two terms of the right hand side of Eq. (\ref{u1}) we 
get,
\bea
W^{\dagger} K(l_{12},m_{12})W \  W^{\dagger}BW  +  W^{\dagger}K(l_{21},m_{21})W \  
W^{\dagger}B^{\dagger}W = l_{12} b_0 W^{\dagger} (B + B^{\dagger}) W \nonumber 
\eea
The expression of $W^{\dagger}(B+B^{\dagger})W$ is,
\bea
W^{\dagger}(B+B^{\dagger})W & = & 2 W_1 W_2 \cos(\theta_1 - \theta_2) \cos^{2}(\epsilon) 
+ \{ 2 W_1 W_2 \big( |\beta|^2 \cos(\theta_1 - \theta_2 + 2 \theta_3) \nonumber \\
                            & - & \Gamma^2 \cos(\theta_1 - \theta_2) \big)  + 
2 \Gamma |\beta| \sin(\theta_3) b_0 \} \ \frac{\sin^2(\epsilon)}{\epsilon^2}
\eea 
The imaginary part of the non linear term vanishes for the choice of $l_{ij}$ and 
$m_{ij}$ as in Eq. (\ref{g4}) and Eq. (\ref{u1}) reduces to the following,
\bea
u_{\zeta\zeta} + \mu u = f(x,t) \ \frac{\rho(-x,t)\rho(x,t)}{\zeta^{2}_{x}} \  
e^{i(\eta(x,t)-\eta(-x,t)} \ u^{*}(-\zeta) u^{2}(\zeta) 
\label{u2}
\eea
and $f(x,t) = \frac{1}{2} [f_1(x,t) + f_2(x,t) + f_3(x,t)]$. 
Note that Eq. (\ref{u2}) reduces to the following equation
\bea
u_{\zeta\zeta} + \mu u - 2 \sigma u^{*}(-\zeta) u^{2}(\zeta) = 0
\label{u3}
\eea
if the following condition is satisfied
\bea
f(x,t) = \frac{2 \sigma \ \zeta^{2}_{x}}{\rho(-x,t)\rho(x,t)} e^{i\big(\eta(-x,t)-\eta(x,t)\big)}
\label{con5}
\eea 
It is to be noted that $f(x,t)$ is a $\cal{PT}$ symmetric function i.e. $f^{*}(-x,t) = f(x,t)$, 
since $\zeta(x,t)$ is an odd function of space. Hence $\zeta_{x}(x,t)$ is an even function of space i.e. 
$\zeta_{x}(-x,t) = \zeta_{x}(x,t)$. The general solution of Eq. (\ref{nlse1}) can be written as,
\bea
\psi(x,t) = U(t) \ W \ \rho(x,t) \ u(\zeta) \ e^{i \eta(x,t)}
\label{final}
\eea
\noindent where $U(t)$ is given in Eq. (\ref{punitary}) and $\rho(x,t)$, $\eta(x,t)$, $\zeta(x,t)$ 
and $u(\zeta)$ are to be determined from Eqs. (\ref{con1}), (\ref{con2}), (\ref{con3}), (\ref{u3}) 
respectively. Eq. (\ref{u3}) is solvable and has many standard solutions. 
The solution of Eq. (\ref{u3}) is as follows,
\begin{eqnarray}
u(\zeta) = \sqrt{d_1} \ sn\{ \sqrt{d_2 |\sigma|} (\zeta - \zeta_0), k \}  
\end{eqnarray}
where we have considered $\sigma < 0$, $\mu > 0$ and $u(\zeta)$ is an odd function. 
Here the constants $d_1$ and $d_2$ are expressed in terms of constants $m,\sigma, c_1$ as shown below,
\bea
d_1 & = & \frac{\mu}{2|\sigma|} - \sqrt{(\frac{\mu}{2|\sigma|})^{2}- \frac{c_1}{\sigma}} \nonumber \\
d_2 & = & \frac{\mu}{2|\sigma|} + \sqrt{(\frac{\mu}{2|\sigma|})^{2}- \frac{c_1}{\sigma}} \nonumber \nonumber
\eea

The expressions of power $ P = \Psi^{\dagger} \Psi$ is given as,

\bea
P(x,t) = P_1(t) \rho^{2}(x,t) u^{2}(\zeta), \ P_1 \equiv W^{\dagger}U^{\dagger}(t)U(t)W \nonumber   
\eea

The time dependent function $P_1(t)$ is determined using the non-unitary matrix $U(t)$ and the expression is given below,
\bea
P_1(t) = b_0  \ [ 1+ \frac{2 \Gamma D}{{\epsilon_0}^{2}} \sin^2(\epsilon) + \frac{b_3 \Gamma}{b_0 \epsilon_0} \sin(2\epsilon) ]
\eea
The expression of $P_1(t)$ is independent of local and nonlocal nature of NLSE. This expression is 
the signature of non-unitary transformation in the expression of power or final solution. 

Different Physically motivated examples are presented in Ref.\cite{pkg2} without balanced loss-gain. 
Those are also applicable here for $\rho(x,t)$, $\zeta(x,t)$, $\eta(x,t)$ for different 
kinds of potentials and $f(x,t)$. The complete solution of Eq. (\ref{nlse1}) is obtained by using Eq. (\ref{final}). 
Some of the results are presented here in the following tables
\begin{table}[h!]
\caption{Examples}
\centering
\begin{tabular}{|c|c|c|c|c|}
\hline
\hline
  & V(x) & f(x) & $\zeta(x)$  & Parameters \\
\hline
\multirow{2}{0.6cm}{Sol-I} & \multirow{2}{2cm}{$E^2 x^2$} & \multirow{2}{2.1cm}{$2 \sigma \ e^{3Ex^2}$} 
& $\sqrt{\frac{\pi}{4E}} \times $ & $E > 0, \mu = 0$ \\
 & & & $erf(\sqrt{E}x)$  & $ \sigma < 0$ \\
\hline
\multirow{2}{0.6cm}{Sol-II} & $N^2 - N(N+1)$ & \multirow{2}{2.1cm}{$2 \sigma \cosh^{6N}(x)$} & 
\multirow{2}{2.6cm}{$\int^{x} \cosh^{2N}(x) dx $} &  $ \mu = 0 , E = 0 $ \\
 &$\times sech^2(x)$ & & & $ \sigma < 0$ \\
\hline
\multirow{2}{0.6cm}{Sol-III} & \multirow{2}{1cm}{0} & \multirow{2}{1.9cm}{$\frac{2\sigma}{\{1+\alpha \cos(\omega x)\}^{3}}$} 
& $\frac{2}{\omega \sqrt{1-\alpha^2}} \times \ $ & $E = \frac{\omega^2}{4} > 0$ \\
 & & & $\arctan[\frac{\sqrt{1-\alpha}}{\sqrt{1+\alpha}} \tan(\frac{\omega x}{2})]$ & $\mu = (1-\alpha^2)E$ \\
\hline
\end{tabular}
\label{table1}
\end{table}
\\
{\bf{\underline{Sol-I}}:} \\
\bea
\psi_1(x,t) & =      & [W_1 e^{i\theta_1} \cos(\epsilon) - \frac{iW_2\beta^{*}}{\epsilon_0} e^{i\theta_2} 
\sin(\epsilon) + \frac{\Gamma W_1}{\epsilon_0} e^{i\theta_1} \sin(\epsilon)] \nonumber \\
            & \times & e^{-\frac{1}{2} E x^2} \frac{c}{\sqrt{2|\sigma|}} \ cn(c\zeta, \frac{1}{2}) \ e^{-iEt} \nonumber \\
\psi_2(x,t) & = & [W_2 e^{i\theta_2} \cos(\epsilon) - \frac{iW_1\beta}{\epsilon_0} e^{i\theta_1} \sin(\epsilon) 
- \frac{\Gamma W_2}{\epsilon_0} e^{i\theta_2} \sin(\epsilon)] \nonumber \\
            & \times & e^{-\frac{1}{2} E x^2} \frac{c}{\sqrt{2|\sigma|}} \ cn(c\zeta, \frac{1}{2}) \ e^{-iEt} 
\eea
{\bf{\underline{Sol-II}}:} \\
\bea
\psi_1(x,t) & =      &  [W_1 e^{i\theta_1} \cos(\epsilon) - \frac{iW_2\beta^{*}}{\epsilon_0} e^{i\theta_2} 
\sin(\epsilon) + \frac{\Gamma W_1}{\epsilon_0} e^{i\theta_1} \sin(\epsilon)]\nonumber \\
            & \times & \sech^{N}(x) \ \frac{c}{\sqrt{2|\sigma|}} \ cn(c \zeta , \frac{1}{2}) \nonumber \\            
\psi_2(x,t) & =      & [W_2 e^{i\theta_2} \cos(\epsilon) - \frac{iW_1\beta}{\epsilon_0} e^{i\theta_1} \sin(\epsilon) 
- \frac{\Gamma W_2}{\epsilon_0} e^{i\theta_2} \sin(\epsilon)] \nonumber \\
            & \times & \sech^{N}(x) \ \frac{c}{\sqrt{2|\sigma|}} \ cn(c \zeta , \frac{1}{2})
\eea 
{\bf{\underline{Sol-III}}:} \\
\bea
\psi_1(x,t) & =      &  [W_1 e^{i\theta_1} \cos(\epsilon) - \frac{iW_2\beta^{*}}{\epsilon_0} e^{i\theta_2} \sin(\epsilon) 
+ \frac{\Gamma W_1}{\epsilon_0} e^{i\theta_1} \sin(\epsilon)]\nonumber \\
            & \times & \sqrt{1 + \alpha \cos(\omega x)} \ \frac{c}{\sqrt{2|\sigma|}} \sqrt{m} 
\ sn(c \zeta ,m) \ e^{-i E t} \ \nonumber \\            
\psi_2(x,t) & =      & [W_2 e^{i\theta_2} \cos(\epsilon) - \frac{iW_1\beta}{\epsilon_0} e^{i\theta_1} 
\sin(\epsilon) - \frac{\Gamma W_2}{\epsilon_0} e^{i\theta_2} \sin(\epsilon)] \nonumber \\
            & \times & \sqrt{1 + \alpha \cos(\omega x)} \ \frac{c}{\sqrt{2|\sigma|}} \sqrt{m} \ sn(c \zeta ,m) \ e^{-i E t} 
\eea 
For Sol-III $\mu = (1+m) c^2$ , $|\alpha| < 1$ and $\sigma < 0$. 

All the above solutions are finite in all regions of space. Sol-I and Sol-II are localized in space as 
$\vert x \vert \rightarrow \infty$, amplitude of the solutions tends to zero. With appropriate choice of the 
parameters, the elliptical functions reduces to hyperbolic function. In this limit, Sol-III also becomes localized.
The amplitude of these solutions oscillates with time. 

\section{Results and Discussions}

We have investigated solvable limits of a class of VNNLSE with time dependent BLG and LC terms 
and space-time modulated nonlinear interaction in presence of confining complex potential. It has been
shown that the system admits Lagrangian and Hamiltonian density in a certain limit. In general, the
Lagrangian density ${\cal{L}}$ is non-hermitian. Further, it is not equal to its complex conjugate
followed by a parity transformation, i. e. ${\cal{L}}(x,t) \neq  {\cal{L}}^{\dagger}(-x,t)$,
as is the case with the NNLSE or its multi-component and higher dimensional generalizations. The
presence of loss-gain terms without the matrix $A$ being $M$-pseudo-hermitian and/or space-time dependence
of $M$ is the reason for ${\cal{L}}(x,t) \neq {\cal{L}}^{\dagger}(-x,t)$. 
The subtleties involved in deriving the Euler-Lagrange equations of motion and conserved Noether
charges associated with invariance under continuous symmetry have been discussed.
Further, the Hamiltonian, charge, width of the wave packet and its speed of the growth of the system
have been shown to be real valued, despite the fact that the corresponding Hamiltonian density,
charge density, current density are complex valued. One necessary condition for these dynamical
variables  to be real valued is that the confining complex potential is $\cal{PT}$-symmetric.
The charge has been shown to be a conserved quantity. We have also presented two constants of
motion in addition to the Hamiltonian and the charge.

The time-evolution of the system has been studied in terms of certain moments which are analogues
of space-integrals of Stokes variables. The VNNLSE can be transformed into a set of linear coupled differential
equations satisfied by these moments provided $A$ is $M$-pseudo-hermitian. The resulting equations can be
solved exactly if the LC and BLG terms have identical time-modulation. The general method presented in this
article for finding solvable limits also requires that the time-modulation of BLG and LC terms are the same.
The regions in the parameter-space for bounded and unbounded solutions in time have been identified
for time-independent BLG and LC terms. It has been shown that with appropriate choice of time-modulation
function $\mu(t)$, the instability in the moments can be tamed. Further, the time-dependence can be
tailor-made by suitably choosing $\mu(t)$. This freedom may be utilized in realistic application of
the system. 

The moment method does not give any expression for the field as a function of space and time.
We have adopted a two step approach to find the exact solutions. In the first step, the BLG and
LC terms are removed completely by a non-unitary transformation which, in general, modifies
the time-modulation of the nonlinear strength. However, in the limit of the loss-gain matrix
$A$ being $M$-pseudo-hermitian and the nonlinear interaction is of Manakov-type,
i.e. ${\vert \Psi^{\dagger}(-x.t) M \Psi(x,t)\vert}^2$, the nonlinear term remains invarinat
under the transformation. The transformation for this case is identified as pseudo-unitary
transformation which is not a symmetry transformation, since it does not preserve the norm. The
resulting VNNLSE can be cast into the canonical form of sovable Manakov-type non-local NLSE through an
$SU(2)$ rotation. The exact solutions of this equation has been used to find the exact solutions of
the system through inverse mapping. Several solutions have been presented for vanishing complex
potential. Further, exact solutions of the system are presented for generalized Rosen-Morse,
Scarf-II and a complex periodic potential.

The next step is required only if the BLG and LC terms are completely removed by imparting
additional time-dependence to the nonlinear strength. The resulting equation for such cases
has been mapped to a solvable equation via a co-ordinate transformation.
The transformed spatial co-ordinate is necessarily odd under parity transformation. This is
to be contrasted with a similar situation in the case of local NLSE where such restriction
on the co-ordinate transformation is not required. Several exact solutions have been found.
The exact solutions do not depend on specific choices of $f_1(x,t)$, $f_2(x,t)$, $f_3(x,t)$ which
appear in the strength of the nonlinear strength, rather depends only on $f(x)=\frac{1}{2} \left
[ f_1(x_)+ f_2(x)+f_3(x) \right ]$. Several choices of $f_i(x)$'s are allowed for a fixed
$f(x)$. Further, the exact solution does not depend on the arbitrary function $G(x,t)$ appearing in the 
nonlinear term. Such a behaviour has been observed for the case of local NLSE also and
the possible reason behind this may be attributed to the specific ansatz. The VNNLSE is characterized
by the functions $f_i(x), i=1, 2, 3$ and  $G(x,t)$ which appear in its nonlinear strength. The
fact that the class of solutions presented in this article does not depend on specific form
of these functions allows to construct a large number of solvable systems for a fixed $f(x)$.
It is desirable that one or more of such solvable VNNLSE may find applications in realistic
physical problems. Further, it is expected that the mapping involving pseudo-unitary transformation
shall be useful for other nonlinear equations too, albeit with appropriate modifications. This
necessitates further investigations involving a variety of nonlinear equations. 

\section{acknowledgements}
SG acknowledges the support of DST INSPIRE fellowship of Govt.of India(Inspire Code No. IF190276). 

\section{Appendix-I}

We have presented two solutions of Eq. (\ref{nlse1}) with $V(x,t) = 0$ and specific choice of $\mu(t)$ 
in Sec-3.1. The basic method involves mapping Eq. (\ref{nlse1}) to Eq. (\ref{p-herm3}) whose solutions
have been discussed in Ref. \cite{Khare}. In this appendix, we present other possible solutions of
Eq. (\ref{nlse1}) with $V(x,t) = 0$ and any arbitrary $\mu(t)$.  It should be noted that all the solutions
of Ref. \cite{Khare} can not be mapped to be exact solutions of Eq. (\ref{p-herm3}), since the
parameters $\lambda_{\pm}$ appearing in the latter equation have specific forms and do not satisfy
the required conditions. For example, There is no solution of Eq. (\ref{p-herm3}) with $V(x,t)=0$
corresponding to the Solution IV and V in Sec-3 of  Ref. \cite{Khare}. Nevertheless, a large number of solutions 
can be found. The solutions presented below do not exhaust all the solutions of  Ref. \cite{Khare} which 
can be mapped to be an exact solutions of Eq. (\ref{p-herm3}), but the complete set of solutions may be found easily.

\noindent \underline{\bf{Solution-III}}

\noindent The solution of Eq. (\ref{nlse1}) with $V(x,t) = 0$ and any arbitrary $\mu(t)$, corresponding 
to the Solution-I in Ref. \cite{Khare} is,

\bea
\psi_1(x,t) & = & \frac{1}{\sqrt{2}} \ \big[ \{ cos(\epsilon) +
(\Gamma - i\beta^{*} e^{i\theta_{\alpha}}) \frac{\sin(\epsilon)}{\epsilon_0} \}
A dn(\Omega x,m) e^{-i (\omega_1 t + \delta_1)} \nonumber \\
            & + & \{-e^{-i\theta_{\alpha}} \cos(\epsilon) -
(\Gamma e^{-i\theta_{\alpha}} + i \beta^{*}) \frac{\sin(\epsilon)}{\epsilon_0} \}
B \sqrt{m} sn(\Omega x,m) e^{-i(\omega_2 t + \delta_2 )}  \big] \nonumber \\
\psi_2(x,t) & = & \frac{1}{\sqrt{2}} \ \big[ \{ \cos(\epsilon) -
(\Gamma - i\beta e^{-i\theta_{\alpha}}) \frac{\sin(\epsilon)}{\epsilon_0} \}
B \sqrt{m} sn(\Omega x,m) e^{-i(\omega_2 t + \delta_2 )} \nonumber \\
            & + & \{e^{i\theta_{\alpha}} \cos(\epsilon) - (\Gamma e^{i\theta_{\alpha}}
+ i \beta) \frac{\sin(\epsilon)}{\epsilon_0} \} A dn(\Omega x,m) e^{-i (\omega_1 t + \delta_1)} \big]
\eea
\noindent provided 
\bea
&& \lambda_{+} A^2 + \lambda_{-} B^2 = - 2 \Omega^2 
\label{rel-1} \\
&& \omega_1 = m \Omega^2 + \lambda_{+} A^2 \ , \ \omega_2 = (1+m) \Omega^2 + \lambda_{+} A^2 
\eea

\noindent \underline{\bf{Solution-IV}}

\noindent The solution of Eq. (\ref{nlse1}) with $V(x,t) = 0$ and any arbitrary $\mu(t)$, corresponding
to the Solution-II in Ref. \cite{Khare} is,

\bea
\psi_1(x,t) & = & \frac{1}{\sqrt{2}} \ \big[ \{ cos(\epsilon) +
(\Gamma - i\beta^{*} e^{i\theta_{\alpha}}) \frac{\sin(\epsilon)}{\epsilon_0} \}
A dn(\Omega x,m) e^{-i (\omega_1 t + \delta_1)} \nonumber \\
            & + & \{-e^{-i\theta_{\alpha}} \cos(\epsilon) -
(\Gamma e^{-i\theta_{\alpha}} + i \beta^{*}) \frac{\sin(\epsilon)}{\epsilon_0} \}
B \sqrt{m} cn(\Omega x,m) e^{-i(\omega_2 t + \delta_2 )}  \big] \nonumber \\
\psi_2(x,t) & = & \frac{1}{\sqrt{2}} \ \big[ \{ \cos(\epsilon) -
(\Gamma - i\beta e^{-i\theta_{\alpha}}) \frac{\sin(\epsilon)}{\epsilon_0} \}
B \sqrt{m} cn(\Omega x,m) e^{-i(\omega_2 t + \delta_2 )} \nonumber \\
            & + & \{e^{i\theta_{\alpha}} \cos(\epsilon) - (\Gamma e^{i\theta_{\alpha}}
+ i \beta) \frac{\sin(\epsilon)}{\epsilon_0} \} A dn(\Omega x,m) e^{-i (\omega_1 t + \delta_1)} \big]
\eea
\noindent provided the parameters satisfy Eq. (\ref{rel-1}) and
\bea
\omega_1 & = & -(4-3m) \Omega^2 + (1-m) \lambda_{+} A^2 \nonumber \\
\omega_2 & = & -(2m-1) \Omega^2 + (1-m) \lambda_{+} A^2
\eea

\noindent \underline{\bf{Solution-V}}

\noindent The solution of Eq. (\ref{nlse1}) with $V(x,t) = 0$ and any arbitrary $\mu(t)$, corresponding
to the Solution-III in Ref. \cite{Khare} is,

\bea
\psi_1(x,t) & = & \frac{1}{\sqrt{2}} \ \big[ \{ cos(\epsilon) +
(\Gamma - i\beta^{*} e^{i\theta_{\alpha}}) \frac{\sin(\epsilon)}{\epsilon_0} \}
\{A dn(\Omega x,m) \nonumber \\
	    & + & D \sqrt{m} cn(\Omega x,m) \} e^{-i (\omega_1 t + \delta_1)} \nonumber \\
            & + & \{-e^{-i\theta_{\alpha}} \cos(\epsilon) -
(\Gamma e^{-i\theta_{\alpha}} + i \beta^{*}) \frac{\sin(\epsilon)}{\epsilon_0} \}
\{B dn(\Omega x,m) \nonumber \\
	    & + & E \sqrt{m} cn(\Omega x,m) \} e^{-i(\omega_2 t + \delta_2 )}  \big] \nonumber \\
\psi_2(x,t) & = & \frac{1}{\sqrt{2}} \ \big[ \{ \cos(\epsilon) -
(\Gamma - i\beta e^{-i\theta_{\alpha}}) \frac{\sin(\epsilon)}{\epsilon_0} \}
\{B dn(\Omega x,m) \nonumber \\
	    & + & E \sqrt{m} cn(\Omega x,m) \} e^{-i(\omega_2 t + \delta_2 )} \nonumber \\
            & + & \{e^{i\theta_{\alpha}} \cos(\epsilon) - (\Gamma e^{i\theta_{\alpha}}
+ i \beta) \frac{\sin(\epsilon)}{\epsilon_0} \} \{A dn(\Omega x,m) \nonumber \\
	    & + & D \sqrt{m} cn(\Omega x,m) \} e^{-i (\omega_1 t + \delta_1)} \big]
\eea
\noindent provided the parameters satisfy Eq. (\ref{rel-1}) and
\bea
\omega_1 = \omega_2 = - \frac{\Omega^2}{2} (1+m) \ , \ D = \pm A \ , \ E = \pm B 
\eea

\noindent \underline{\bf{Solution-VI}}

\noindent The solution of Eq. (\ref{nlse1}) with $V(x,t) = 0$ and any arbitrary $\mu(t)$, corresponding
to the Solution-VII in Ref. \cite{Khare} is,

\bea
\psi_1(x,t) & = & \frac{1}{\sqrt{2}} \ \big[ \{ cos(\epsilon) +
(\Gamma - i\beta^{*} e^{i\theta_{\alpha}}) \frac{\sin(\epsilon)}{\epsilon_0} \}
A \sqrt{m} \frac{cn(\Omega x,m)}{dn(\Omega x,m)} e^{-i (\omega_1 t + \delta_1)} \nonumber \\
            & + & \{-e^{-i\theta_{\alpha}} \cos(\epsilon) -
(\Gamma e^{-i\theta_{\alpha}} + i \beta^{*}) \frac{\sin(\epsilon)}{\epsilon_0} \}
B \sqrt{m(1-m)} \frac{sn(\Omega x,m)}{dn(\Omega x,m)} e^{-i(\omega_2 t + \delta_2 )}  \big] \nonumber \\
\psi_2(x,t) & = & \frac{1}{\sqrt{2}} \ \big[ \{ \cos(\epsilon) -
(\Gamma - i\beta e^{-i\theta_{\alpha}}) \frac{\sin(\epsilon)}{\epsilon_0} \}
B \sqrt{m(1-m)} \frac{sn(\Omega x,m)}{dn(\Omega x,m)} e^{-i(\omega_2 t + \delta_2 )} \nonumber \\
            & + & \{e^{i\theta_{\alpha}} \cos(\epsilon) - (\Gamma e^{i\theta_{\alpha}}
+ i \beta) \frac{\sin(\epsilon)}{\epsilon_0} \} A \sqrt{m} \frac{cn(\Omega x,m)}{dn(\Omega x,m)} 
e^{-i (\omega_1 t + \delta_1)} \big]
\eea
\noindent provided 
\bea
&& \lambda_{+} A^2 + \lambda{-} B^2 = 2 \Omega^2 \nonumber \\
&& \omega_1  =  (1-m) \Omega^2 + m \lambda_{+} A^2 \nonumber \\
&& \omega_2  =  (1-2m) \Omega^2 + m \lambda_{+} A^2
\eea

\noindent \underline{\bf{Solution-VII}}

\noindent The solution of Eq. (\ref{nlse1}) with $V(x,t) = 0$ and any arbitrary $\mu(t)$, corresponding
to the Solution-VIII in Ref. \cite{Khare} is,

\bea
\psi_1(x,t) & = & \frac{1}{\sqrt{2}} \ \big[ \{ cos(\epsilon) +
(\Gamma - i\beta^{*} e^{i\theta_{\alpha}}) \frac{\sin(\epsilon)}{\epsilon_0} \}
A \sqrt{m} \frac{cn(\Omega x,m)}{dn^2(\Omega x,m)} e^{-i (\omega_1 t + \delta_1)} \nonumber \\
            & + & \{-e^{-i\theta_{\alpha}} \cos(\epsilon) -
(\Gamma e^{-i\theta_{\alpha}} + i \beta^{*}) \frac{\sin(\epsilon)}{\epsilon_0} \}
B m \frac{cn(\Omega x,m) sn(\Omega x,m)}{dn^2(\Omega x,m)} e^{-i(\omega_2 t + \delta_2 )}  \big] \nonumber \\
\psi_2(x,t) & = & \frac{1}{\sqrt{2}} \ \big[ \{ \cos(\epsilon) -
(\Gamma - i\beta e^{-i\theta_{\alpha}}) \frac{\sin(\epsilon)}{\epsilon_0} \}
B m \frac{cn(\Omega x,m) sn(\Omega x,m)}{dn^2(\Omega x,m)} e^{-i(\omega_2 t + \delta_2 )} \nonumber \\
            & + & \{e^{i\theta_{\alpha}} \cos(\epsilon) - (\Gamma e^{i\theta_{\alpha}}
+ i \beta) \frac{\sin(\epsilon)}{\epsilon_0} \} A \sqrt{m} \frac{cn(\Omega x,m)}{dn^2(\Omega x,m)}
e^{-i (\omega_1 t + \delta_1)} \big]
\eea
\noindent provided 
\bea
&& \omega_1 = (4m+1) \Omega^2 \ , \ \omega_2 = (4+m) \Omega^2 \nonumber \\
&& \lambda_{+} A^2 = \lambda_{-} B^2 = 6 \Omega^2 
\eea

\noindent \underline{\bf{Solution-VIII}} 

\noindent The solution of Eq. (\ref{nlse1}) with $V(x,t) = 0$ and any arbitrary $\mu(t)$, corresponding
to the Solution-IX in Ref. \cite{Khare} is,

\bea
\psi_1(x,t) & = & \frac{1}{\sqrt{2}} \ \big[ \{ cos(\epsilon) +
(\Gamma - i\beta^{*} e^{i\theta_{\alpha}}) \frac{\sin(\epsilon)}{\epsilon_0} \}
A \sqrt{m} cn(\Omega x,m) e^{-i (\omega_1 t + \delta_1)} \nonumber \\
            & + & \{-e^{-i\theta_{\alpha}} \cos(\epsilon) -
(\Gamma e^{-i\theta_{\alpha}} + i \beta^{*}) \frac{\sin(\epsilon)}{\epsilon_0} \}
B \sqrt{m} sn(\Omega x,m) e^{-i(\omega_2 t + \delta_2 )}  \big] \nonumber \\
\psi_2(x,t) & = & \frac{1}{\sqrt{2}} \ \big[ \{ \cos(\epsilon) -
(\Gamma - i\beta e^{-i\theta_{\alpha}}) \frac{\sin(\epsilon)}{\epsilon_0} \}
B \sqrt{m} sn(\Omega x,m) e^{-i(\omega_2 t + \delta_2 )} \nonumber \\
            & + & \{e^{i\theta_{\alpha}} \cos(\epsilon) - (\Gamma e^{i\theta_{\alpha}}
+ i \beta) \frac{\sin(\epsilon)}{\epsilon_0} \} A \sqrt{m} cn(\Omega x,m)
e^{-i (\omega_1 t + \delta_1)} \big]
\eea
\noindent provided the parameters satisfy Eq. (\ref{rel-1}) and
\bea
\omega_1 = \Omega^2 + m \lambda_{+} A^2 \ , \ \omega_2 = m \Omega^2 + \omega_1
\eea

\noindent \underline{\bf{Solution-IX}} 

\noindent The solution of Eq. (\ref{nlse1}) with $V(x,t) = 0$ and any arbitrary $\mu(t)$, corresponding
to the Solution-X in Ref. \cite{Khare} is,

\bea
\psi_1(x,t) & = & \frac{1}{\sqrt{2}} \ \big[ \{ cos(\epsilon) +
(\Gamma - i\beta^{*} e^{i\theta_{\alpha}}) \frac{\sin(\epsilon)}{\epsilon_0} \}
A \sqrt{m} sn(\Omega x,m) e^{-i (\omega_1 t + \delta_1)} \nonumber \\
            & + & \{-e^{-i\theta_{\alpha}} \cos(\epsilon) -
(\Gamma e^{-i\theta_{\alpha}} + i \beta^{*}) \frac{\sin(\epsilon)}{\epsilon_0} \}
B \sqrt{m} sn(\Omega x,m) e^{-i(\omega_2 t + \delta_2 )}  \big] \nonumber \\
\psi_2(x,t) & = & \frac{1}{\sqrt{2}} \ \big[ \{ \cos(\epsilon) -
(\Gamma - i\beta e^{-i\theta_{\alpha}}) \frac{\sin(\epsilon)}{\epsilon_0} \}
B \sqrt{m} sn(\Omega x,m) e^{-i(\omega_2 t + \delta_2 )} \nonumber \\
            & + & \{e^{i\theta_{\alpha}} \cos(\epsilon) - (\Gamma e^{i\theta_{\alpha}}
+ i \beta) \frac{\sin(\epsilon)}{\epsilon_0} \} A \sqrt{m} sn(\Omega x,m)
e^{-i (\omega_1 t + \delta_1)} \big]
\eea
\noindent provided the parameters satisfy Eq. (\ref{rel-1}) and
\bea
\omega_1 = \omega_2 = (1+m) \Omega^2  
\eea

\noindent \underline{\bf{Solution-X}} 

\noindent The solution of Eq. (\ref{nlse1}) with $V(x,t) = 0$ and any arbitrary $\mu(t)$, corresponding
to the Solution-XI in Ref. \cite{Khare} is,

\bea
\psi_1(x,t) & = & \frac{1}{\sqrt{2}} \ \big[ \{ cos(\epsilon) +
(\Gamma - i\beta^{*} e^{i\theta_{\alpha}}) \frac{\sin(\epsilon)}{\epsilon_0} \}
A \sqrt{m} cn(\Omega x,m) e^{-i (\omega_1 t + \delta_1)} \nonumber \\
            & + & \{-e^{-i\theta_{\alpha}} \cos(\epsilon) -
(\Gamma e^{-i\theta_{\alpha}} + i \beta^{*}) \frac{\sin(\epsilon)}{\epsilon_0} \}
B \sqrt{m} cn(\Omega x,m) e^{-i(\omega_2 t + \delta_2 )}  \big] \nonumber \\
\psi_2(x,t) & = & \frac{1}{\sqrt{2}} \ \big[ \{ \cos(\epsilon) -
(\Gamma - i\beta e^{-i\theta_{\alpha}}) \frac{\sin(\epsilon)}{\epsilon_0} \}
B \sqrt{m} cn(\Omega x,m) e^{-i(\omega_2 t + \delta_2 )} \nonumber \\
            & + & \{e^{i\theta_{\alpha}} \cos(\epsilon) - (\Gamma e^{i\theta_{\alpha}}
+ i \beta) \frac{\sin(\epsilon)}{\epsilon_0} \} A \sqrt{m} cn(\Omega x,m)
e^{-i (\omega_1 t + \delta_1)} \big]
\eea
\noindent provided the parameters satisfy Eq. (\ref{rel-1}) and
\bea
\omega_1 = \omega_2 = -(2m-1) \Omega^2 
\eea

\noindent \underline{\bf{Solution-XI}} 

\noindent The solution of Eq. (\ref{nlse1}) with $V(x,t) = 0$ and any arbitrary $\mu(t)$, corresponding
to the Solution-XII in Ref. \cite{Khare} is,

\bea
\psi_1(x,t) & = & \frac{1}{\sqrt{2}} \ \big[ \{ cos(\epsilon) +
(\Gamma - i\beta^{*} e^{i\theta_{\alpha}}) \frac{\sin(\epsilon)}{\epsilon_0} \}
A \sqrt{m} dn(\Omega x,m) e^{-i (\omega_1 t + \delta_1)} \nonumber \\
            & + & \{-e^{-i\theta_{\alpha}} \cos(\epsilon) -
(\Gamma e^{-i\theta_{\alpha}} + i \beta^{*}) \frac{\sin(\epsilon)}{\epsilon_0} \}
B \sqrt{m} dn(\Omega x,m) e^{-i(\omega_2 t + \delta_2 )}  \big] \nonumber \\
\psi_2(x,t) & = & \frac{1}{\sqrt{2}} \ \big[ \{ \cos(\epsilon) -
(\Gamma - i\beta e^{-i\theta_{\alpha}}) \frac{\sin(\epsilon)}{\epsilon_0} \}
B \sqrt{m} dn(\Omega x,m) e^{-i(\omega_2 t + \delta_2 )} \nonumber \\
            & + & \{e^{i\theta_{\alpha}} \cos(\epsilon) - (\Gamma e^{i\theta_{\alpha}}
+ i \beta) \frac{\sin(\epsilon)}{\epsilon_0} \} A \sqrt{m} dn(\Omega x,m)
e^{-i (\omega_1 t + \delta_1)} \big]
\eea
\noindent provided the parameters satisfy Eq. (\ref{rel-1}) and
\bea
\omega_1 = \omega_2 = -(2-m) \Omega^2 
\eea

\noindent \underline{\bf{Solution-XII}} 

\noindent The solution of Eq. (\ref{nlse1}) with $V(x,t) = 0$ and any arbitrary $\mu(t)$, corresponding
to the Solution-XIV in Ref. \cite{Khare} is,

\bea
\psi_1(x,t) & = & \frac{1}{\sqrt{2}} \ \big[ \{ cos(\epsilon) +
(\Gamma - i\beta^{*} e^{i\theta_{\alpha}}) \frac{\sin(\epsilon)}{\epsilon_0} \}
A \tanh(\Omega x) e^{-i (\omega_1 t + \delta_1)} \nonumber \\
            & + & \{-e^{-i\theta_{\alpha}} \cos(\epsilon) -
(\Gamma e^{-i\theta_{\alpha}} + i \beta^{*}) \frac{\sin(\epsilon)}{\epsilon_0} \}
B \tanh(\Omega x) e^{-i(\omega_2 t + \delta_2 )}  \big] \nonumber \\
\psi_2(x,t) & = & \frac{1}{\sqrt{2}} \ \big[ \{ \cos(\epsilon) -
(\Gamma - i\beta e^{-i\theta_{\alpha}}) \frac{\sin(\epsilon)}{\epsilon_0} \}
B \tanh(\Omega x) e^{-i(\omega_2 t + \delta_2 )} \nonumber \\
            & + & \{e^{i\theta_{\alpha}} \cos(\epsilon) - (\Gamma e^{i\theta_{\alpha}}
+ i \beta) \frac{\sin(\epsilon)}{\epsilon_0} \} A \tanh(\Omega x)
e^{-i (\omega_1 t + \delta_1)} \big]
\eea
\noindent provided the parameters satisfy Eq. (\ref{rel-1}) and
\bea
\omega_1 = \omega_2 = 2 \Omega^2
\eea

\noindent \underline{\bf{Solution-XIV}} 

\noindent The solution of Eq. (\ref{nlse1}) with $V(x,t) = 0$ and any arbitrary $\mu(t)$, corresponding
to the Solution-XV in Ref. \cite{Khare} is,

\bea
\psi_1(x,t) & = & \frac{1}{\sqrt{2}} \ \big[ \{ cos(\epsilon) +
(\Gamma - i\beta^{*} e^{i\theta_{\alpha}}) \frac{\sin(\epsilon)}{\epsilon_0} \}
A \sech(\Omega x) e^{-i (\omega_1 t + \delta_1)} \nonumber \\
            & + & \{-e^{-i\theta_{\alpha}} \cos(\epsilon) -
(\Gamma e^{-i\theta_{\alpha}} + i \beta^{*}) \frac{\sin(\epsilon)}{\epsilon_0} \}
B \sech(\Omega x) e^{-i(\omega_2 t + \delta_2 )}  \big] \nonumber \\
\psi_2(x,t) & = & \frac{1}{\sqrt{2}} \ \big[ \{ \cos(\epsilon) -
(\Gamma - i\beta e^{-i\theta_{\alpha}}) \frac{\sin(\epsilon)}{\epsilon_0} \}
B \sech(\Omega x) e^{-i(\omega_2 t + \delta_2 )} \nonumber \\
            & + & \{e^{i\theta_{\alpha}} \cos(\epsilon) - (\Gamma e^{i\theta_{\alpha}}
+ i \beta) \frac{\sin(\epsilon)}{\epsilon_0} \} A \sech(\Omega x)
e^{-i (\omega_1 t + \delta_1)} \big]
\eea
\noindent provided the parameters satisfy Eq. (\ref{rel-1}) and
\bea
\omega_1 = \omega_2 = -2 \Omega^2
\eea

\noindent \underline{\bf{Solution-XV}} 

\noindent The solution of Eq. (\ref{nlse1}) with $V(x,t) = 0$ and any arbitrary $\mu(t)$, corresponding
to the Solution-XXII in Ref. \cite{Khare} is
\bea
\psi_1(x,t) & = & \frac{1}{\sqrt{2}} \ \big[ \{ cos(\epsilon) +
(\Gamma - i\beta^{*} e^{i\theta_{\alpha}}) \frac{\sin(\epsilon)}{\epsilon_0} \}
\{A \sech^2(\Omega x) + D\} e^{-i (\omega_1 t + \delta_1)} \nonumber \\
            & + & \{-e^{-i\theta_{\alpha}} \cos(\epsilon) -
(\Gamma e^{-i\theta_{\alpha}} + i \beta^{*}) \frac{\sin(\epsilon)}{\epsilon_0} \}
B \tanh(\Omega x) \sech(\Omega x) e^{-i(\omega_2 t + \delta_2 )}  \big] \nonumber \\
\psi_2(x,t) & = & \frac{1}{\sqrt{2}} \ \big[ \{ \cos(\epsilon) -
(\Gamma - i\beta e^{-i\theta_{\alpha}}) \frac{\sin(\epsilon)}{\epsilon_0} \}
B \tanh(\Omega x) \sech(\Omega x) e^{-i(\omega_2 t + \delta_2 )} \nonumber \\
            & + & \{e^{i\theta_{\alpha}} \cos(\epsilon) - (\Gamma e^{i\theta_{\alpha}}
+ i \beta) \frac{\sin(\epsilon)}{\epsilon_0} \} \{A \sech^2(\Omega x) + D \}
e^{-i (\omega_1 t + \delta_1)} \big]
\eea
\noindent provided 
\bea
&& \lambda_{+} A^2 = - \lambda_{-} B^2 = 18 \Omega^2 \nonumber \\
&& \omega_1 = 8 \Omega^2 \ , \ \omega_2 = 7 \Omega^2 \ , \ \frac{D}{A} = - \frac{2}{3}
\eea

\end{document}